\documentclass[11pt]{myArticle}       

\usepackage{amsmath,amssymb,amsfonts} 
\usepackage{graphicx}
\usepackage{subfig}
\usepackage{epsfig}
\usepackage{color}                    
\usepackage{hyperref}                 
\usepackage{myFancyhdr}               
\usepackage{makeidx}                  
\usepackage{helvet}                   
\usepackage{setspace}                 
\usepackage{fancybox}                 
\usepackage{float}                    
\usepackage{afterpage}                
\usepackage{xspace}                   
\usepackage{wasysym}                  
\usepackage[numbers,sort&compress]{natbib}
\usepackage{natbib}
\usepackage{hypernat}
\usepackage{wrapfig}
\usepackage{lineno}
\usepackage{bm}
\usepackage{lscape}                   
\usepackage{tikz}                     
\usepackage{multirow}
\usepackage{textcomp}

\definecolor{mygray}{rgb}{0.3,0.32,0.35}
\definecolor{darkblue1}{rgb}{0,0,.2}
\definecolor{darkblue}{rgb}{0,0,.3}
\definecolor{darkred}{rgb}{0.5,0,0}
\pagecolor{white} 
\color{black}     
\hypersetup{breaklinks=true, 
            colorlinks=true, 
            linkcolor=darkblue1, 
            menucolor=darkblue1, 
            urlcolor=darkblue1,
            citecolor=darkblue1,
            pdftitle={The EW Fit},
            pdfauthor={Gfitter Group},
            pdfsubject={The EW Fit},
            pdfkeywords={},
            pdfproducer={Gfitter Group}
}
\newcommand\defaultSingleFigureScale{0.70}

\newcommand\defaultDoubleFigureScaleTikz{0.48}

\newlength{\captionverticalgap}
\bibstyle{plain}
\renewcommand{\headrulewidth}{0.5pt}    
\renewcommand{\footrulewidth}{0.5pt}    

\newcommand\allFontSize{\small}

\newenvironment{myquote}
               {\list{}{\leftmargin0cm}%
                \item\relax}
               {\endlist}

\newcommand\detailsSize{\allFontSize}
\newenvironment{details}%
{\begin{myquote}\vspace{-0.2cm}\detailsSize}{\end{myquote}\vspace{-0.2cm}}

\newfloat{codeexample}{H}{loc}
\floatname{codeexample}{\sf\allFontSize Code Example}

\newlength{\gfitterboxwidth}
\definecolor{DarkGray}{rgb}{0.4,0.42,0.45}
\definecolor{LightGray}{rgb}{0.97,0.98,0.98}

{\VerbatimEnvironment\normalsize
\setlength{\gfitterboxwidth}{\textwidth}\addtolength{\gfitterboxwidth}{-2.3\fboxsep}%
\begin{Sbox}\begin{minipage}[t]{\gfitterboxwidth}\begin{Verbatim}}%
{\end{Verbatim}\end{minipage}\end{Sbox}\vspace{1ex} \fcolorbox{DarkGray}{LightGray}{\TheSbox}}

\newfloat{option}{H}{loc}
\floatname{option}{\sf\allFontSize Option Table}

\usepackage{tabularx}
{\setlength{\gfitterboxwidth}{\textwidth}\addtolength{\gfitterboxwidth}{-2.3\fboxsep}%
\begin{Sbox}\begin{tabular*}{\gfitterboxwidth}{>{\rule[-0.5ex]{.0ex}{3.0ex}\tt\small}p{3cm}>{\tt\small}p{4.5cm}>{\small}p{5.7cm}}
&&\\[-0.7cm]
\rm\small Option  & \rm\small Values    & \rm\small Description\\[0.1cm]\hline
&&\\[-0.45cm]}%
{\\[-0.1cm]\end{tabular*}\end{Sbox}\vspace{1ex} \fbox{\TheSbox}}

{\setlength{\gfitterboxwidth}{\textwidth}\addtolength{\gfitterboxwidth}{-2.3\fboxsep}%
\begin{Sbox}\begin{tabular*}{\gfitterboxwidth}{>{\rule[-0.5ex]{.0ex}{3.0ex}\tt\small}p{3cm}>{\tt\small}p{3.5cm}>{\small}p{6.7cm}}
&&\\[-0.7cm]
\rm\small Option  & \rm\small Values    & \rm\small Description\\[0.1cm]\hline
&&\\[-0.45cm]}%
{\\[-0.1cm]\end{tabular*}\end{Sbox}\vspace{1ex} \fbox{\TheSbox}}

{\setlength{\gfitterboxwidth}{\textwidth}\addtolength{\gfitterboxwidth}{-2.3\fboxsep}%
\begin{Sbox}\begin{tabular*}{\gfitterboxwidth}{>{\rule[-0.5ex]{.0ex}{3.5ex}\tt\small}p{3.0cm}>{\tt\small}p{3.0cm}>{\small}p{7.2cm}}
&&\\[-0.7cm]
\rm\small Option  & \rm\small Values    & \rm\small Description\\[0.1cm]\hline
&&\\[-0.45cm]}%
{\\[-0.1cm]\end{tabular*}\end{Sbox}\vspace{1ex} \fbox{\TheSbox}}

{\setlength{\gfitterboxwidth}{\textwidth}\addtolength{\gfitterboxwidth}{-2.3\fboxsep}%
\begin{Sbox}\begin{tabular*}{\gfitterboxwidth}{>{\rule[-0.5ex]{.0ex}{3.0ex}\tt\small}p{2.5cm}>{\tt\small}p{3.5cm}>{\small}p{7.2cm}}
&&\\[-0.7cm]
\rm\small Option  & \rm\small Values    & \rm\small Description\\[0.1cm]\hline
&&\\[-0.45cm]}%
{\\[-0.1cm]\end{tabular*}\end{Sbox}\vspace{1ex} \fbox{\TheSbox}}

{\setlength{\gfitterboxwidth}{\textwidth}\addtolength{\gfitterboxwidth}{-2\fboxsep}%
\begin{Sbox}\begin{tabular*}{\gfitterboxwidth}{>{\rule[-0.5ex]{.0ex}{3.0ex}\tt\small}p{4.5cm}>{\tt\small}p{3.2cm}>{\small}p{5.5cm}}
&&\\[-0.7cm]
\rm\small Option  & \rm\small Values    & \rm\small Description\\[0.1cm]\hline
&&\\[-0.45cm]}%
{\\[-0.1cm]\end{tabular*}\end{Sbox}\vspace{1ex} \fbox{\TheSbox}}

{\setlength{\gfitterboxwidth}{\textwidth}\addtolength{\gfitterboxwidth}{-2\fboxsep}%
\begin{Sbox}\begin{tabular*}{\gfitterboxwidth}{>{\rule[-0.5ex]{.0ex}{3.0ex}\tt\small}p{4.0cm}>{\tt\small}p{3.0cm}>{\small}p{6.2cm}}
&&\\[-0.7cm]
\rm\small Option  & \rm\small Values    & \rm\small Description\\[0.1cm]\hline
&&\\[-0.45cm]}%
{\\[-0.1cm]\end{tabular*}\end{Sbox}\vspace{1ex} \fbox{\TheSbox}}

\newfloat{programs}{H}{loc}
\floatname{programs}{\sf Table}

\usepackage{tabularx}
{\setlength{\gfitterboxwidth}{\textwidth}\addtolength{\gfitterboxwidth}{-2\fboxsep}%
\begin{Sbox}\begin{tabular*}{\gfitterboxwidth}{>{\rule[-0.5ex]{.0ex}{3.0ex}\tt\small}p{4.1cm}>{\small}p{9.5cm}}
&\\[-0.7cm]
\rm\small Macro & \rm\small Description\\[0.1cm]\hline
&\\[-0.45cm]}%
{\\[-0.1cm]\end{tabular*}\end{Sbox}\vspace{1ex} \fbox{\TheSbox}}

\newcommand{\ccbar} {\ensuremath{c\overline c}\xspace}

\newcommand{\bbbar} {\ensuremath{b\overline b}\xspace}

\newcommand{\ttbar} {\ensuremath{t\overline t}\xspace}

\newcommand{\MSbar}{\ensuremath{\overline{\rm MS}}\xspace}
\mathchardef\Upsilon="7107
\def\Y#1S{\ensuremath{\Upsilon{(#1S)}}\xspace}

\newcommand{\stat}{\ensuremath{_\mathrm{stat}}\xspace}
\newcommand{\syst}{\ensuremath{_\mathrm{syst}}\xspace}
\newcommand{\kappai}{\ensuremath{\kappa_i}\xspace}
\newcommand{\kappaf}{\ensuremath{\kappa_f}\xspace}
\newcommand{\setkappa}{\ensuremath{\bm{\kappa}}\xspace}
\newcommand{\kappag}{\ensuremath{\kappa_g}\xspace}
\newcommand{\kappaga}{\ensuremath{\kappa_\gamma}\xspace}
\newcommand{\kappaZga}{\ensuremath{\kappa_{Z\gamma}}\xspace}
\newcommand{\kappatau}{\ensuremath{\kappa_{\tau}}\xspace}
\newcommand{\kappamu}{\ensuremath{\kappa_{\mu}}\xspace}
\newcommand{\kappat}{\ensuremath{\kappa_t}\xspace}
\newcommand{\kappab}{\ensuremath{\kappa_b}\xspace}
\newcommand{\kappac}{\ensuremath{\kappa_c}\xspace}
\newcommand{\kappaV}{\ensuremath{\kappa_V}\xspace}
\newcommand{\kappaW}{\ensuremath{\kappa_W}\xspace}
\newcommand{\kappaZ}{\ensuremath{\kappa_Z}\xspace}
\newcommand{\kappaF}{\ensuremath{\kappa_F}\xspace}
\newcommand{\kappaH}{\ensuremath{\kappa_H}\xspace}
\newcommand{\GammaH}{\ensuremath{\Gamma_{H}}\xspace}
\newcommand{\GammaHtot}{\ensuremath{\Gamma_{H,\mathrm{tot}}}\xspace}
\newcommand{\GammaHSM}{\ensuremath{\Gamma_{H,\mathrm{SM}}}\xspace}
\newcommand{\GammaBSM}{\ensuremath{\Gamma_{\mathrm{i+u}}}\xspace}
\newcommand{\BRBSM}{\ensuremath{\BR_{\mathrm{i+u}}}\xspace}
\newcommand{\BRHinv}{\ensuremath{\BR_{\mathrm{inv}}}\xspace}
\newcommand{\BRHundet}{\ensuremath{\BR_{\mathrm{undet}}}\xspace}

\newcommand{\tH}{\ensuremath{tH}\xspace}
\newcommand{\ttH}{\ensuremath{t\bar{t}H}\xspace}
\newcommand{\ttHtH}{\ensuremath{t\bar{t}H + tH}\xspace}
\newcommand{\ggH}{\ensuremath{ggH}\xspace}
\newcommand{\VBF}{\ensuremath{\rm VBF}\xspace}
\newcommand{\VH}{\ensuremath{V\mkern-2mu H}\xspace}
\newcommand{\WH}{\ensuremath{W\mkern-2mu H}\xspace}
\newcommand{\ZH}{\ensuremath{Z\mkern-1mu H}\xspace}
\newcommand{\HVV}{\ensuremath{HV\mkern-1mu V\!}\xspace}

\newcommand{\mt}{\ensuremath{m_{t}}\xspace}
\newcommand{\MW}{\ensuremath{M_{W}}\xspace}
\newcommand{\MH}{\ensuremath{M_{H}}\xspace}
\newcommand{\as}{\ensuremath{\alpha_{\scriptscriptstyle S}}\xspace}

\newcommand{\asZ}{\ensuremath{\as(M_Z^2)}\xspace}

\renewcommand\l{\ell}

\newcommand{\GF}{{\ensuremath{G_{\mkern-1mu\scriptscriptstyle F}}}\xspace}

\newcommand{\Kbar    }{\kern 0.2em\overline{\kern -0.2em K}{}\xspace}

\newcommand{\Kz      }{\ensuremath{K^0}\xspace}
\newcommand{\Kzb     }{\ensuremath{\Kbar^0}\xspace}
\newcommand{\KzKzb   }{\ensuremath{\Kz \kern -0.16em \Kzb}\xspace}
\newcommand{\Kp      }{\ensuremath{K^+}\xspace}
\newcommand{\Km      }{\ensuremath{K^-}\xspace}

\newcommand{\KpKm    }{\ensuremath{\Kp \kern -0.16em \Km}\xspace}

\newcommand\Dbar    {\kern 0.18em\overline{\kern -0.18em D}{}\xspace}

\newcommand\Bbar    {\kern 0.18em\overline{\kern -0.18em B}{}\xspace}

\newcommand\Bz      {\ensuremath{B^0}\xspace}

\newcommand\Bzb     {\ensuremath{\Bbar^0}\xspace}
\newcommand\Bu      {\ensuremath{B^+}\xspace}
\newcommand\Bub     {\ensuremath{B^-}\xspace}

\newcommand\BpBm    {\ensuremath{\Bu {\kern -0.16em \Bub}}\xspace}
\newcommand\Bs      {\ensuremath{B^0_{s}}\xspace}
\newcommand\Bsb     {\ensuremath{\Bbar^0_{s}}\xspace}

\newcommand\BzBzb   {\ensuremath{\Bz {\kern -0.16em \Bzb}}\xspace}
\newcommand\BszBszb {\ensuremath{\Bs {\kern -0.16em \Bsb}}\xspace}

\newcommand\deltatheo{\ensuremath{\delta_{\rm th}}\xspace}
\newcommand\deltaambi{\ensuremath{\delta_{\ambi}}\xspace}

\newcommand{\ft}{\footnotesize}

\newcommand{\Order}{\ensuremath{{\cal O}}\xspace}
\newcommand{\BR}{\ensuremath{{\mathcal B}}\xspace}

\newcommand{\ee}{\ensuremath{e^+e^-}\xspace}
\newcommand{\mm}{\ensuremath{\mu^+\mu^-}\xspace}

\newcommand{\tev}{\ensuremath{\mathrm{\:Te\kern -0.1em V}}\xspace}
\newcommand{\gev}{\ensuremath{\mathrm{\:Ge\kern -0.1em V}}\xspace}
\newcommand{\mev}{\ensuremath{\mathrm{\:Me\kern -0.1em V}}\xspace}
\newcommand{\kev}{\ensuremath{\mathrm{\:ke\kern -0.1em V}}\xspace}
\newcommand{\ev}{\ensuremath{\mathrm{\,e\kern -0.1em V}}\xspace}
\newcommand{\gevc}{\ensuremath{{\mathrm{\:Ge\kern -0.1em V\!/}c}}\xspace}
\newcommand{\mevc}{\ensuremath{{\mathrm{\:Me\kern -0.1em V\!/}c}}\xspace}
\newcommand{\gevcc}{\ensuremath{{\mathrm{\:Ge\kern -0.1em V\!/}c^2}}\xspace}
\newcommand{\mevcc}{\ensuremath{{\mathrm{\:Me\kern -0.1em V\!/}c^2}}\xspace}

\newcommand{\bei}{\begin{itemize}}
\newcommand{\eei}{\end{itemize}}
\newcommand{\beq}{\begin{equation}}
\newcommand{\eeq}{\end{equation}}
\newcommand{\beqn}{\begin{eqnarray}}
\newcommand{\eeqn}{\end{eqnarray}}
\newcommand{\beqns}{\begin{eqnarray*}}
\newcommand{\eeqns}{\end{eqnarray*}}
\newcommand{\bitm}{\begin{itemize}}
\newcommand{\eitm}{\end{itemize}}

\newcommand{\dalphaHadMZ}{\ensuremath{\Delta\alpha_{\rm had}^{(5)}(M_Z^2)}\xspace}

\newcommand\ie{{i.e.}\xspace}

\newcommand\cf{{cf.}\xspace}

\newcommand\rs{\raisebox{1.5ex}[-1.5ex]}

\def\@citex[#1]#2{\if@filesw\immediate\write\@auxout{\string\citation{#2}}\fi
  \@tempcnta\z@\@tempcntb\m@ne\def\@citea{}\@cite{\@for\@citeb:=#2\do
    {\@ifundefined
       {b@\@citeb}{\@citeo\@tempcntb\m@ne\@citea
        \def\@citea{,\penalty\@m\ }{\bf ?}\@warning
       {Citation `\@citeb' on page \thepage \space undefined}}%
    {\setbox\z@\hbox{\global\@tempcntc0\csname b@\@citeb\endcsname\relax}%
     \ifnum\@tempcntc=\z@ \@citeo\@tempcntb\m@ne
       \@citea\def\@citea{,\penalty\@m}
       \hbox{\csname b@\@citeb\endcsname}%
     \else
      \advance\@tempcntb\@ne
      \ifnum\@tempcntb=\@tempcntc
      \else\advance\@tempcntb\m@ne\@citeo
      \@tempcnta\@tempcntc\@tempcntb\@tempcntc\fi\fi}}\@citeo}{#1}}

\def\@citeo{\ifnum\@tempcnta>\@tempcntb\else\@citea
  \def\@citea{,\penalty\@m}%
  \ifnum\@tempcnta=\@tempcntb\the\@tempcnta\else
   {\advance\@tempcnta\@ne\ifnum\@tempcnta=\@tempcntb \else
\def\@citea{--}\fi
    \advance\@tempcnta\m@ne\the\@tempcnta\@citea\the\@tempcntb}\fi\fi}

\xspace

\newcommand\experi{{\rm exp}}
\newcommand\ambi{{\rm amb}}

\newcommand\ChiMin{\ensuremath{\chi^2_{\min}}\xspace}
\newcommand\Ndof{\ensuremath{N_{\mathrm{dof}}}\xspace}
\newcommand\DeltaChi{\ensuremath{\Delta\chi^2}\xspace}

\newcommand{\seffsf}[1]{\sin\!^2\theta^{#1}_{{\rm eff}}}

\newcommand{\sinfeff}{\ensuremath{\seffsf{f}}\xspace}
\newcommand{\sinbeff}{\ensuremath{\seffsf{b}}\xspace}
\newcommand{\sinleff}{\ensuremath{\seffsf{\ell}}\xspace}

\newcommand{\mc}{\ensuremath{\overline{m}_c}\xspace}
\newcommand{\mb}{\ensuremath{\overline{m}_b}\xspace}

\newcommand{\SParam}     	{\ensuremath{0.03\pm 0.10}\xspace}
\newcommand{\TParam}     	{\ensuremath{0.05\pm 0.12}\xspace}
\newcommand{\UParam}     	{\ensuremath{-0.03\pm 0.09}\xspace}

\newcommand{\STParamCor}	{\ensuremath{+0.92}\xspace}
\newcommand{\SUParamCor}	{\ensuremath{-0.67}\xspace}
\newcommand{\TUParamCor}	{\ensuremath{-0.88}\xspace}

\newcommand{\SParamNU}     {\ensuremath{0.01\pm 0.07}\xspace}
\newcommand{\TParamNU}     {\ensuremath{0.02\pm 0.05}\xspace}
\newcommand{\STParamCorNo} {\ensuremath{+0.94}\xspace}

\newcommand{\kappaVEWPO}   {\ensuremath{1.009\,^{+0.012}_{-0.010}}\xspace}
\newcommand{\GammaHHVV}    {\ensuremath{4.36\,^{+0.63}_{-0.50}\mev}\xspace}
\newcommand{\GammaHEff}    {\ensuremath{4.08\,^{+0.43}_{-0.37}\mev}\xspace}
\newcommand{\GammaHRes}    {\ensuremath{3.78\,^{+0.30}_{-0.27}\mev}\xspace}

\newcommand{\BRBSMVV}      {\ensuremath{0.27}\xspace}
\newcommand{\BRBSMgen}     {\ensuremath{0.17}\xspace}
\newcommand{\BRBSMres}     {\ensuremath{0.09}\xspace}
\newcommand{\BRBSMnoEWPO}  {\ensuremath{0.07}\xspace}

\makeindex
\begin{document}


%
%
\pagenumbering{arabic}
{\small
\color{mygray}
\begin{flushright}
{\sf\em DESY-26-089} \\
{\sf\em \today} \\
\def\UrlFont{\sf\em}
\url{http://cern.ch/gfitter} 
\end{flushright}
}
\def\UrlFont{\rm}

\vspace{1.3cm}


{\sf\LARGE\bfseries
The Higgs boson through the lens of electroweak \\[0.15cm] 
precision data
}

\vspace{1.0cm}

{\large \em 
  The Gfitter Group \\[0.2cm]
}
{\large
  Y.~Fischer$^{a}$, J.~Haller$^{a}$, A.~Hoecker$^{b}$, R.~Kogler$^{c}$, F.~Labe$^{c}$, 
  K.~M\"onig$^{c}$, D.~Schwarz$^{c}$, J.~Stelzer$^{d}$
}

\vspace{-0.2cm}

\begin{details}
  $^{a}$Institut f\"ur Experimentalphysik, Universit\"at Hamburg, Germany\\
  $^{b}$CERN, Geneva, Switzerland \\
  $^{c}$Deutsches Elektronen-Synchrotron DESY, Germany \\ 
  $^{d}$University of Pittsburgh, PA, USA 
\end{details}

\vspace{1.0cm}

\begin{details} \centerline{\footnotesize\bf Abstract} 
The global electroweak fit tests the quantum structure of the Standard Model
by confronting precision measurements with high-order theoretical predictions.
This paper presents an updated Gfitter analysis using the latest experimental inputs, 
notably the new world average of the $W$-boson mass, and state-of-the-art theoretical 
calculations. The fit yields indirect determinations of precision observables, including 
the $W$ and Higgs-boson masses, the top-quark mass, and the effective leptonic weak 
mixing angle, confirming the remarkable internal consistency of the Standard Model. 
The analysis is further extended to the Higgs sector by combining ATLAS and CMS 
signal-strength measurements with electroweak precision data in the $\kappa$ 
framework. The resulting electroweak constraints on the Higgs couplings to vector 
bosons allow a determination of the total Higgs-boson width with a precision of 
about 10\% or better within a leading-logarithmic oblique interpretation, and provide 
bounds on invisible and undetected Higgs-boson branching fractions without using 
direct searches for invisible Higgs decays. The paper also presents bounds on Wilson 
coefficients in the Standard Model Effective Field Theory together with projections 
for the precision of the global electroweak fit achievable at the FCC-ee.

\end{details}

\thispagestyle{empty}

\newpage
{\small\setlength{\parskip}{3pt}
\renewcommand{\baselinestretch}{0.88}\selectfont
\tableofcontents
}\pagebreak

%
%
%

\section{Introduction}
\label{sec:intro}

Precision measurements of the electroweak sector have provided some of the most 
incisive tests of the Standard Model for decades. Results from LEP, SLC, the 
Tevatron and the LHC~\cite{ALEPH:2005ema,Schael:2013ita,TevatronElectroweakWorkingGroup:2012gb,Aaltonen:2018dxj,ATLAS:2024wla,ATLAS:2024kxj,CMS:2024gzs,Aaij:2015lka,LHCb:2024ygc,CERNCourier:GoKr} probe 
the theory through quantum corrections, giving 
indirect access to its fundamental parameters. 

This sensitivity gave the global electroweak fit a predictive power~\cite{Ellis:1988ap,Langacker:1991an,Langacker:1991zr,Ellis:1992qs,Blondel:1993jq,Ellis:1994se,ALEPH:2005ema,Erler:2010wa,Flacher:2008zq,Baak:2011ze} 
that was spectacularly confirmed by the subsequent discoveries of the top 
quark~\cite{CDF:1995wbb,D0:1995jca} and the Higgs boson~\cite{Aad:2012tfa,Chatrchyan:2012xdj}. 
With these discoveries and the corresponding mass measurements, the SM electroweak 
fit became fully constrained~\cite{Baak:2014ora}. It has since evolved from a tool of indirect 
prediction into a stringent test of the internal consistency of the SM and a 
sensitive probe of possible contributions from physics beyond the SM (BSM).

Global electroweak fits are performed by several groups and frameworks, including 
Gfitter~\cite{Flacher:2008zq,Baak:2011ze,Baak:2012kk,Baak:2014ora,Haller:2018nnx, Erler:2019hds}, 
GAPP~\cite{Erler:1994fz,Erler:1999ug,Erler:2019hds}, and HEPfit~\cite{DeBlas:2019ehy,
deBlas:2021wap,deBlas:2022hdk}, as well as by the LEP Electroweak Working 
Group~\cite{ALEPH:2005ema,Schael:2013ita}, which uses the ZFITTER software~\cite{Bardin:1997xq,Bardin:1999yd,Arbuzov:2005ma}. By confronting SM predictions with a broad 
set of precision observables, these fits test the consistency of the electroweak 
sector at the loop level and constrain possible BSM effects.

The extensive programme of Higgs-boson property measurements at the LHC has 
opened a complementary direction for precision tests of the SM. Measurements 
of Higgs-boson production and decay rates probe its couplings to known particles 
and provide sensitivity to its total width. They also constrain possible decays 
into invisible or undetected final states that may arise in BSM scenarios. 
Electroweak precision data add important information to this programme, 
since modifications of the Higgs-boson couplings to electroweak gauge bosons 
would affect quantum corrections to precision observables. Their combination 
with Higgs signal-strength measurements therefore provides a powerful way 
to test the Higgs sector beyond the direct information contained in LHC rate 
measurements alone.

This paper presents an updated global electroweak fit using the Gfitter framework, 
incorporating recent experimental inputs and state-of-the-art predictions 
for electroweak precision observables. The update includes, among others, the new 
$W$-boson mass combination~\cite{MWaverage:May2026} and recent LHC measurements 
of the effective leptonic weak mixing angle~\cite{CMS:2024ony, LHCb:2024ygc}. 
Updated indirect determinations 
are reported for all observables, together with revised constraints on the 
oblique parameters, which parametrise possible BSM loop 
contributions.

The analysis is then extended to the Higgs sector. Electroweak precision data 
are used to constrain the Higgs-boson couplings to electroweak gauge bosons 
through their contributions to the oblique parameters~\cite{Peskin:1990zt,Peskin:1991sw}. 
These constraints are combined with the latest ATLAS and CMS signal-strength 
combinations to perform Higgs-coupling fits under several sets of assumptions. 
The combined analysis provides a determination of the total Higgs-boson width 
and constraints on the branching fraction to the sum of invisible and undetectable 
Higgs-boson decays.

Constraints on Wilson coefficients of dimension-six Standard Model Effective 
Field Theory (SMEFT) operators are also derived from electroweak precision data. 
Results obtained with fixed SM input parameters are compared with fits in which the 
SM parameters and Wilson coefficients are determined simultaneously.

Finally, prospects for the Future Circular $e^+e^-$ Collider 
(FCC-ee)~\cite{deBlas:2025gyz} are studied using projected uncertainties for 
electroweak and Higgs-boson observables. The expected precision is compared with 
the current state of the art, illustrating the potential of the FCC-ee to sharpen the 
global consistency tests of the SM and to extend their sensitivity to BSM effects.

The paper is organised as follows. Section~\ref{sec:ewfit} describes the updated
global electroweak fit. Section~\ref{sec:higgs} presents the Higgs-coupling fits,
including the width determination and constraints on invisible and undetected
branching fractions. Section~\ref{sec:smeft} gives the SMEFT interpretation.
Section~\ref{sec:fccee} presents the FCC-ee projections, followed by conclusions
in Section~\ref{sec:conclusions}.

\section{The global electroweak fit}
\label{sec:ewfit}

The global electroweak fit combines the most precise measurements of electroweak 
precision observables (EWPO) with state-of-the-art SM predictions. The fit presented in 
this section is performed with the Gfitter framework. Details of the statistical 
procedure are given in our previous publications~\cite{Flacher:2008zq,Baak:2011ze,
Baak:2012kk,Baak:2014ora,Haller:2018nnx}.

Since our last publication~\cite{Haller:2018nnx}, several new measurements of 
EWPO and improved theoretical calculations have become 
available. While no single update substantially changes the fit, their combination 
in a simultaneous analysis leads to reduced uncertainties compared with our previous 
results, enabling a test of the SM with unprecedented precision.

The observables used in the fit, together with their values and uncertainties, are 
described below. 

\subsection{Experimental inputs}
\label{sec:impr}

The experimental inputs comprise the masses and widths of the $Z$ and $W$ bosons, 
the Higgs-boson mass, the top-quark mass, the effective leptonic weak mixing angle, 
the strong coupling constant, the hadronic contribution to the running  
electromagnetic coupling, and the $Z$-pole observables measured at LEP and SLC. 
A complete list of observables, including their values and uncertainties, is 
given in the first two columns of Table~\ref{tab:results}. Below, we summarise 
the experimental inputs, focusing on quantities updated with respect to our 
previous analysis~\cite{Haller:2018nnx}. For a recent detailed review of the 
measurements, we refer to Ref.~\cite{EW-ParticleDataGroup:2026}.

\subsubsection*{Higgs-boson mass}

The Higgs-boson mass has been measured with high precision at the LHC. We use 
the average~\cite{ParticleDataGroup:2026}
\beq
M_H = 125.13 \pm 0.11\gev,
\eeq
which includes the ATLAS and CMS Run-1 combination~\cite{ATLAS:2015yey}, the ATLAS 
Run-2 combination~\cite{ATLAS:2023oaq}, and the CMS Run-2 measurements in the 
$H\to\gamma\gamma$ and $H\to ZZ^{*}\to 4\ell$ channels~\cite{CMS:2020xrn,CMS:2024eka}. 
The quoted uncertainty includes a scale factor that accounts for the 
mild tension between the CMS $H\to\gamma\gamma$ result~\cite{CMS:2020xrn} and 
the other inputs. Its impact on the fit is negligible.

\subsubsection*{{\em Z} and {\em W}-boson masses and widths}

The most precise determination of the $Z$-boson mass remains the LEP combination of the 
ALEPH, DELPHI, L3 and OPAL measurements, $M_Z = 91.1876 \pm 0.0021\gev$~\cite{ALEPH:2005ema}, 
whose uncertainty is dominated by the calibration of the LEP beam energy. More 
recent hadron-collider measurements, obtained by CDF in dimuon and dielectron final 
states~\cite{CDF:2022hxs} and by LHCb using the dimuon decay~\cite{LHCb:2025nob}, 
are combined with the LEP result to obtain the world average~\cite{ParticleDataGroup:2026}
\beq
M_Z = 91.1879 \pm 0.0020\gev,
\eeq
which is used in the fit. The recent CMS determination~\cite{CMS:2024lrd}, 
obtained as a by-product of its $W$-mass measurement, is not included pending 
a dedicated assessment of systematic uncertainties. The total $Z$-boson width is taken 
from the LEP combination, $\Gamma_Z = 2.4955 \pm 0.0023\gev$~\cite{ALEPH:2005ema}, 
updated for the improved Bhabha cross section~\cite{Janot:2019oyi}.

\begin{figure}[tb]
  {\centering
    \includegraphics[width=0.6\textwidth]{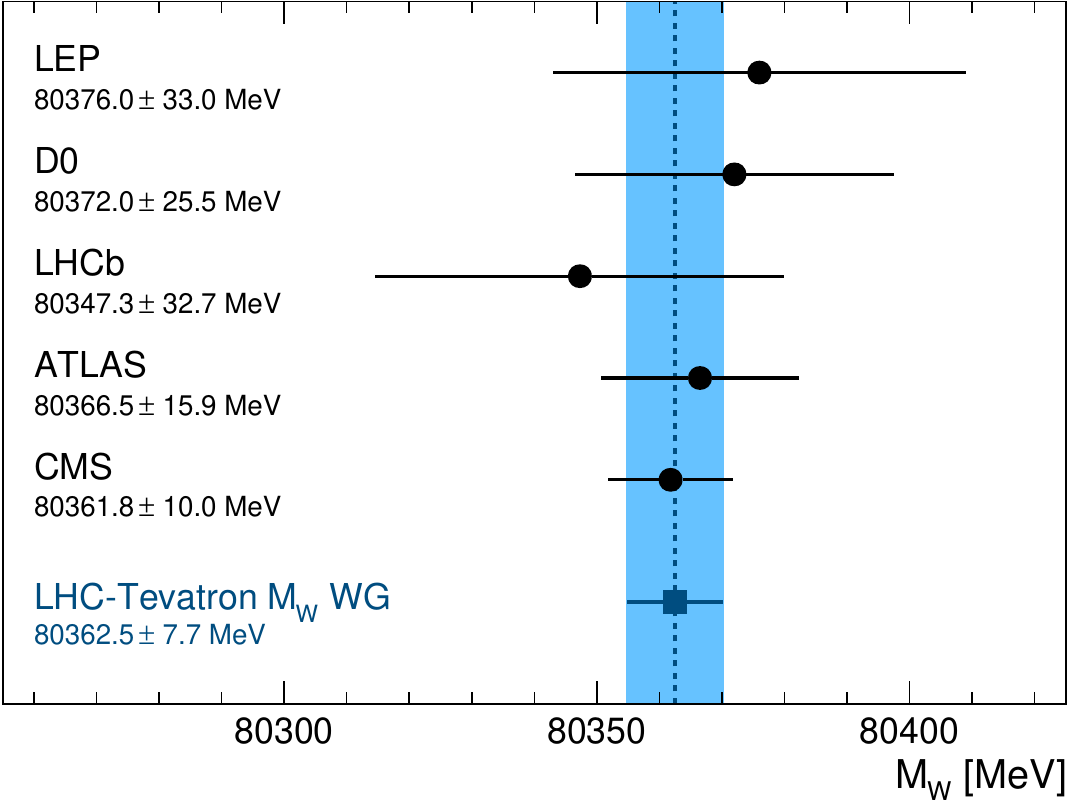} \\
  }
  \vspace{\captionverticalgap}
  \caption[]{Measurements of \MW entering the combination by the LHC-Tevatron $M_W$
  Working Group~\cite{MWaverage:May2026}, together with the resulting average. 
  \label{fig:mw}}  
\end{figure}

For the $W$-boson mass, we use the May 2026 update of the LHC-Tevatron $M_W$ 
Working Group combination~\cite{MWaverage:May2026}. The combination follows the 
methodology of Ref.~\cite{LHC-TeVMWWorkingGroup:2023zkn} and includes measurements 
from ATLAS~\cite{ATLAS:2024erm}, CMS~\cite{CMS:2024lrd}, LHCb~\cite{LHCb:2021bjt}, 
D0~\cite{D0:2012kms,D0:2013jba}, and LEP~\cite{Schael:2013ita}. All measurements are 
evaluated using the CT18 PDF set~\cite{Hou:2019efy} and are shown in Fig.~\ref{fig:mw}. 
Their uncertainties are decomposed into elementary components~\cite{Pinto:2023yob}, and the PDF-induced correlations are propagated coherently across experiments. The resulting average has a $p$-value of $0.97$, 
\beq
\MW = 80.3625 \pm 0.0077\gev,
\eeq
and is used in the fit. The CDF measurement~\cite{CDF:2022hxs}, 
$\MW = 80.432 \pm 0.016\gev$ after adjustment to the CT18 PDF set, differs by almost 
$4\sigma$ from the other measurements and is not included. Its inclusion would reduce 
the $p$-value of the combination to about $6\cdot10^{-4}$.

The $W$-boson width was measured for the first time at the LHC by ATLAS yielding 
$\Gamma_W = 2.202 \pm 0.047\gev$~\cite{ATLAS:2024erm}, currently the single most 
precise determination of this quantity. It is consistent with the LEP 
measurement, $\Gamma_W = 2.195 \pm 0.083\gev$~\cite{Schael:2013ita}, but differs by 
about two standard deviations from the Tevatron combination, 
$\Gamma_W = 2.046 \pm 0.049\gev$~\cite{FERMILAB-TM-2460-E}. We use the weighted average 
of these three measurements, $\Gamma_W = 2.14 \pm 0.05\gev$~\cite{ParticleDataGroup:2026}, 
where the uncertainty has been scaled by a factor of 1.7 to account for the low 
combination $p$-value of 0.05.

\subsubsection*{{\em Z}-pole observables}

A large set of $Z$-pole observables was measured at LEP and SLC and combined by the 
LEP Electroweak Working Group and the SLD Collaboration~\cite{ALEPH:2005ema}.\footnote{Lepton universality is assumed throughout this article.} Besides $M_Z$ and $\Gamma_Z$, discussed 
above, these include the ratios of partial decay widths
$R_\ell^0 \equiv \Gamma_{\rm had}/\Gamma_{\ell\ell}= 20.767 \pm 0.025$,
$R_b^0 \equiv \Gamma_{b\bar b}/\Gamma_{\rm had}= 0.21629 \pm 0.00066$, and
$R_c^0 \equiv \Gamma_{c\bar c}/\Gamma_{\rm had}= 0.1721 \pm 0.0030$.
The hadronic peak cross section is $\sigma_{\rm had}^0 = 41.480 \pm 0.033~\mathrm{nb}$, 
where the value has been updated for the improved Bhabha cross section~\cite{Janot:2019oyi} 
and for a newer calculation of beam-beam effects relevant to the LEP luminosity 
measurement~\cite{Voutsinas:2019hwu}. Among the $Z$-pole observables, $R_\ell^0$ 
and $\Gamma_Z$ provide the dominant sensitivity to the strong coupling 
constant.

The angular distributions of the produced fermions in $Z$ decays give access to the 
effective weak couplings and, hence, to the effective weak mixing angle. We include 
the forward-backward asymmetries measured at LEP,
$A_{\rm FB}^{0,\ell} = 0.0171 \pm 0.0010$ for leptons, and
$A_{\rm FB}^{0,b} = 0.0996 \pm 0.0016$ and
$A_{\rm FB}^{0,c} = 0.0707 \pm 0.0035$ for $b$ and $c$ quarks, together with the 
effective weak mixing angle extracted directly from the hadronic charge asymmetry,
$\sinleff(Q_{\rm FB}) = 0.2324 \pm 0.0012$.
The value of $A_{\rm FB}^{0,b}$ differs from that in Ref.~\cite{ALEPH:2005ema} 
because of the $m_b$ dependence~\cite{Bernreuther:2016ccf} of the two-loop QCD 
correction~\cite{Catani:1999nf,Djouadi:1989uk}.

Exploiting the longitudinal beam polarisation at SLC, SLD determined the asymmetry 
parameters $A_f$ directly from left-right asymmetries. The resulting leptonic 
asymmetry, $A_\ell = 0.1513 \pm 0.0021$~\cite{ALEPH:2005ema}, is more precise 
than the corresponding LEP measurement, $A_\ell = 0.1465 \pm 0.0033$~\cite{ALEPH:2005ema}. 
We use these two measurements as separate inputs to the fit, together with 
the heavy-quark asymmetries
$A_b = 0.923 \pm 0.020$,
$A_c = 0.670 \pm 0.027$, and
$A_s = 0.895 \pm 0.091$~\cite{Abe:2000dq,Abe:2000uc,Abe:2000hk}.
The correlations between the extracted $Z$-pole observables~\cite{ALEPH:2005ema}
are included in the fit.

\subsubsection*{Effective leptonic weak mixing angle from hadron colliders}

The effective leptonic weak mixing angle, $\sinleff$, has been measured at 
hadron colliders through the forward-backward asymmetry in Drell–Yan lepton-pair 
production. The Tevatron combination of the CDF~\cite{Aaltonen:2016nuy} and 
D0~\cite{Abazov:2017gpw} measurements gives $\sinleff = 0.23148 \pm 0.00033~$~\cite{Aaltonen:2018dxj}. At the LHC, measurements are available from ATLAS at $7\tev$~\cite{Aad:2015uau} and 
$8\tev$~\cite{ATLAS-CONF-2018-037}, CMS at $8\tev$~\cite{Sirunyan:2018swq} and 
$13\tev$~\cite{CMS:2024ony}, and LHCb at $7$ and $8\tev$~\cite{Aaij:2015lka} and 
$13\tev$~\cite{LHCb:2024ygc}.
\begin{figure}
  {\centering
    \includegraphics[width=0.55\textwidth]{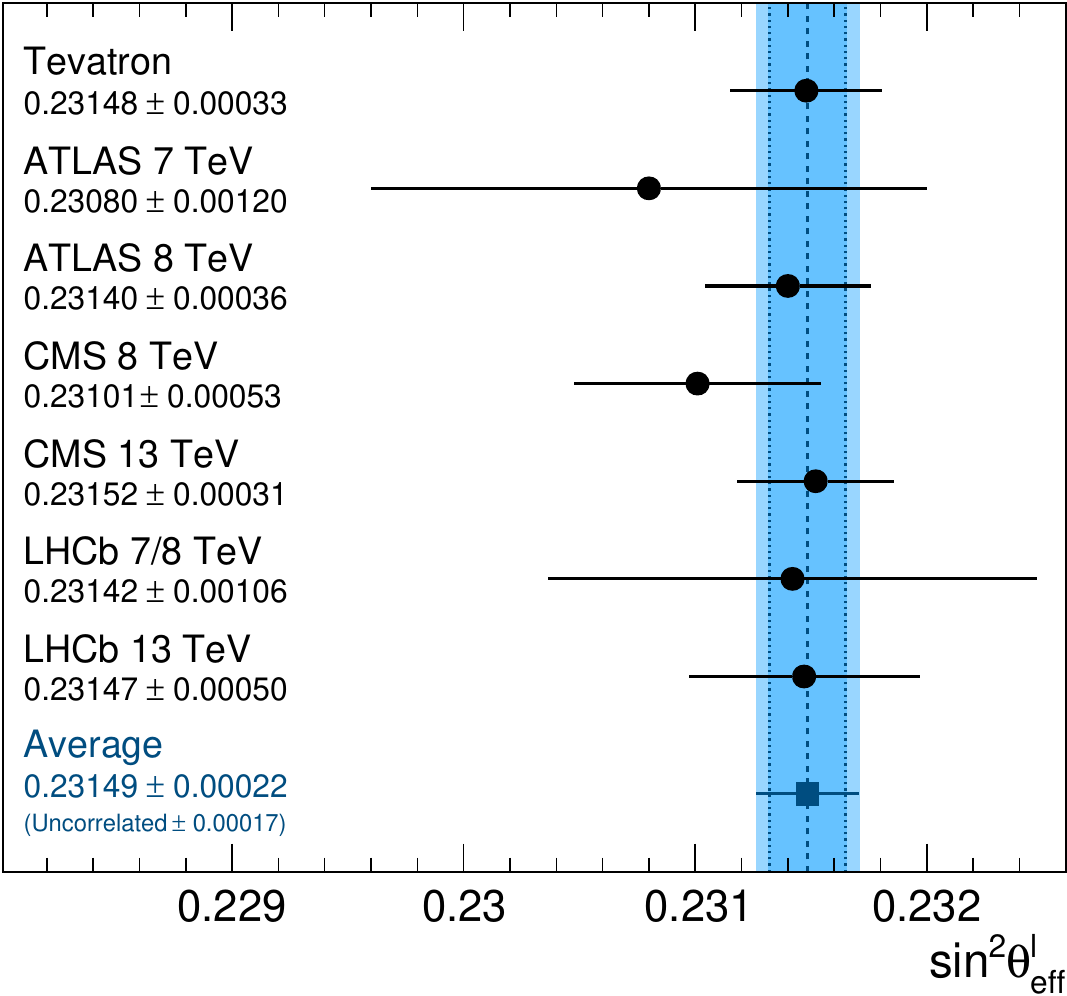} \\
  }
  \vspace{\captionverticalgap}
  \caption[]{Hadron-collider measurements of \sinleff entering our average, 
  together with the combined value. The outer band of the combined value 
  shows the total uncertainty of $\pm 0.00021$, while the inner band shows, 
  for comparison, an uncertainty of $\pm 0.00017$, obtained for fully uncorrelated 
  uncertainties of the individual measurements. 
  \label{fig:sin}
  }
\end{figure}
In the absence of a combination by the experimental collaborations, we combine 
these measurements into a single fit input. Since their dominant systematic 
uncertainty arises from PDFs, assumptions on PDF correlations are required. 
The measurements generally use different PDF sets, except for the Tevatron 
combination and the CMS $8\tev$ result. We do not correct for these differences, 
but assign partial PDF correlations: 25\% between the Tevatron and LHC measurements, 
reflecting the small overlap in parton momentum fraction; 25\% between LHCb and 
ATLAS/CMS; 75\% between measurements at $7$ and $8\tev$; and 50\% between Run~1 
at $7$ and $8\tev$, and Run~2 at $13\tev$. Statistical and experimental 
uncertainties are treated as uncorrelated between experiments, while experimental 
uncertainties from the same experiment at different centre-of-mass energies 
are taken to be 50\% correlated.

With these assumptions, we obtain the hadron-collider (HC) combination
\beq
\sinleff(\mathrm{HC}) = 0.23149 \pm 0.00021\,,
\label{eq:sinefflep_hc}
\eeq
with a $p$-value of 0.97. The contributing measurements and average are 
shown in Fig.~\ref{fig:sin}.

This average is kept separate from the \sinleff\ information already contained 
in the LEP and SLD $Z$-pole asymmetry measurements. For comparison, these measurements 
give $\sinleff = 0.23150 \pm 0.00016$. Although their uncertainty remains smaller 
than that of the HC combination, the latter measurements have become competitive 
with the precision reached at lepton colliders.

To assess the sensitivity to the assumed correlations, we vary the PDF correlations 
by $\pm25\%$. The upward variation is combined with a 50\% upward variation of the 
experimental correlations, while the downward variation assumes fully uncorrelated 
experimental uncertainties. The resulting shift in the central value, $\pm0.00006$, 
is included as an additional uncertainty in the total uncertainty quoted in 
Eq.~\eqref{eq:sinefflep_hc}. If all uncertainties are instead treated as fully 
uncorrelated, the total uncertainty of the HC combination is $\pm0.00017$.

\subsubsection*{Quark masses}

The top quark induces the largest fermionic radiative 
corrections to EWPO, making \mt\ an essential input to the fit. Direct measurements, 
based on the reconstruction of top-quark decay products, are available from the 
Tevatron combination, $\mt = 174.30 \pm 0.35\stat \pm 0.54\syst\gev$~\cite{TevatronElectroweakWorkingGroup:2016lid}, and the ATLAS and CMS Run-1 
combination $\mt = 172.52 \pm 0.14\stat \pm 0.30\syst\gev$~\cite{CMS:2023wnd}. 
These are complemented by several Run-2 measurements from 
ATLAS~\cite{ATLAS:2022jbw,ATLAS:2025bpp,ATLAS:2022jpn} in the lepton+jets, 
dilepton and all-jets channels, as well as from single-top-quark production,
and from CMS~\cite{CMS:2023ebf,CMS:2022kqg,Sirunyan:2018goh,CMS:2018tye,CMS:2021jnp}
summarised in Ref.~\cite{CMS:2024irj}.

Because of the tension between the Tevatron and LHC Run-1 combinations, we do 
not combine them, but use the more recent LHC Run-1 result in the fit. A theoretical 
uncertainty of $0.5\gev$ is added to account for the ambiguity in relating the 
directly measured mass to the top-quark pole mass entering the radiative 
corrections~\cite{Hoang:2020iah,Hoang:2018zrp,Marquard:2015qpa,Beneke:2016cbu,
Beneke:1998ui}, yielding
\beq
\mt = 172.52 \pm 0.33_{\experi} \pm 0.50_{\ambi}\gev\,.
\eeq
This precision still exceeds that of theoretically cleaner pole-mass 
determinations from inclusive and differential \ttbar production cross 
sections~\cite{ATLAS:2014nxi,ATLAS:2017dhr,ATLAS:2025iza,CMS:2014rml,CMS:2016yys,
CMS:2017xrt}, which nevertheless offer promising prospects for future precision measurements.

The charm and bottom-quark masses play a minor role in the fit, entering mainly 
through the running of the electromagnetic coupling and the hadronic $Z$-pole 
observables. We use the running \MSbar\ masses evaluated at their own 
scale, $\mc = 1.273 \pm 0.003\gev$ and $\mb = 4.183 \pm 0.004\gev$~\cite{Erler:2022mzd}.

\subsubsection*{Strong coupling constant}

Although the strong coupling constant does not enter the EWPO at leading order, 
it contributes through radiative corrections to the hadronic $Z$-pole 
observables at one loop, and to \MW and \sinleff from two loops onward. 
The most stringent global fit is obtained by including external determinations 
of $\asZ$. Following the procedure of Ref.~\cite{EW-ParticleDataGroup:2026}, we 
average the value obtained without input from the electroweak fit, 
$\asZ = 0.1171 \pm 0.0011$, with the FLAG2024 lattice-QCD result, 
$\asZ = 0.1183 \pm 0.0007$~\cite{FlavourLatticeAveragingGroupFLAG:2024oxs}, 
giving
\beq
\asZ = 0.1177 \pm 0.0009\,,
\label{eq:alphas_pdg}
\eeq
which is used in the fit. Previously, no external constraint on \asZ 
was used in the fit, instead, its value was determined in situ. However, 
the new precise measurements of $M_W$ at the LHC challenge the theoretical predictions 
and therefore necessitate the smallest possible parametric uncertainties.

\subsubsection*{Hadronic contribution to the electromagnetic coupling}

The electromagnetic coupling at the $Z$-mass scale, $\alpha(M_Z^2)$, enters the prediction 
of all EWPO. Its leptonic contribution is known with high precision in perturbation theory, 
whereas the hadronic contribution from the five light quarks, \dalphaHadMZ, cannot 
be computed perturbatively. It is instead obtained via the optical theorem and 
analyticity from a dispersion integral over the $e^+e^-\to\mathrm{hadrons}$ cross 
section, with experimental data used at low energies and near quark-antiquark 
thresholds, and perturbative QCD at intermediate and high energies.

We use the evaluation of Ref.~\cite{Davier:2019can},
\beq
\dalphaHadMZ = (2758 \pm 10)\cdot 10^{-5}\,,
\eeq
with the central value shifted by $-2\cdot 10^{-5}$ to account for the value of 
\asZ in Eq.~\eqref{eq:alphas_pdg}. This dependence is taken into account in the 
electroweak fit. Discrepancies among low-energy cross-section 
measurements have led to tensions in the dispersion integral evaluations, 
and also with lattice QCD calculations, of the hadronic contribution to the 
anomalous magnetic moment of the muon~\cite{Davier:2023fpl,Aliberti:2025beg}. 
Owing to the different kernel function, 
the impact of these discrepancies is less pronounced for \dalphaHadMZ. We point out,
however, that replacing the current two-pion contribution used in Ref.~\cite{Davier:2019can} 
by one based on the CMD-3 data alone~\cite{CMD-3:2023alj,CMD-3:2023rfe} in the corresponding 
energy region would increase\footnote{Using the 
value $\dalphaHadMZ = (2769 \pm 10)\cdot 10^{-5}$ shifts the indirect 
predictions of $M_H$ by $-5\gev$, $m_t$ by $+0.4\gev$, $M_W$ by $-1.6\mev$, and 
$\sinleff$ by $+3.6\cdot10^{-5}$. The $\chi^2$ value of the global fit increases 
from 13.8 to 14.3.} \dalphaHadMZ by $11\cdot 10^{-5}$ compared to the 
results reported in Table~\ref{tab:results}.

The top-quark contribution to $\alpha(M_Z^2)$, which depends on \mt, is computed 
perturbatively and included separately in the theoretical predictions of the EWPO.

\subsection{Theoretical calculations}
\label{sec:theory}

The SM predictions used in the fit include the most precise higher-order radiative 
corrections available in the literature. They are implemented in the on-shell 
renormalisation scheme through semi-analytical parametrisations. We summarise 
their current status below.

For the effective weak mixing angle, we use the parametrisations of 
Refs.~\cite{Dubovyk:2019szj,Awramik:2006ar}. They include the complete electroweak 
two-loop corrections~\cite{Awramik:2004ge,Awramik:2006uz,Hollik:2005va,Awramik:2006ar,
Hollik:2006ma,Awramik:2008gi,Dubovyk:2016aqv,Dubovyk:2018rlg}, including the bosonic 
two-loop corrections to the $Z\bbbar$ vertex~\cite{Dubovyk:2016aqv}, as well as partial 
higher-order QCD corrections of order $\alpha\alpha_s^2$ and $\alpha\alpha_s^3$, and 
the leading top-mass-enhanced terms of order $\alpha^2\alpha_sm_t^4$ and 
$\alpha^3 m_t^6$~\cite{Avdeev:1994db,Chetyrkin:1995ix,Chetyrkin:1995js,vanderBij:2000cg,
Faisst:2003px,Schroder:2005db,Chetyrkin:2006bj,Boughezal:2006xk}. The predictions for 
charged leptons and $b$ quarks are taken from Ref.~\cite{Dubovyk:2019szj}, while 
those for the other fermions follow Ref.~\cite{Awramik:2006ar}. More recent three-loop 
electroweak corrections involving closed fermion loops, of order $\alpha^3$ and 
$\alpha^2\alpha_s$~\cite{Chen:2020xzx,Chen:2020xot}, as well as subleading 
$\mathcal{O}(\alpha\alpha_s)$ contributions~\cite{Kniehl:1988ie,Halzen:1990je,Djouadi:1993ss,Dittmaier:2020vra}, are numerically small. The bosonic two-loop corrections to 
the $Z\bbbar$ vertex shift the prediction of $A_{\rm FB}^{0,b}$ by $1.3\cdot10^{-5}$, 
two orders of magnitude below the experimental uncertainty, and therefore do not 
affect the fit results.

The partial and total $Z$-boson widths and the hadronic peak cross section, 
$\sigma^0_{\rm had}$, are computed including the complete fermionic electroweak 
two-loop corrections and the leading bosonic two-loop contributions~\cite{Freitas:2013dpa,Freitas:2014hra,Freitas:2012sy,Dubovyk:2018rlg,Dubovyk:2019szj,Chen:2022dow}. 
The dominant effects from final-state QED and QCD radiation are also 
included~\cite{Kataev:1992dg,Czarnecki:1996ei,Harlander:1997zb,Kniehl:1988ie,
Kniehl:1989qu,Chetyrkin:1993yp,Larin:1993ju,Chetyrkin:1993ug,Chetyrkin:1994js,
Baikov:2008jh,Baikov:2012er}. We use the parametrisations of Ref.~\cite{Dubovyk:2019szj}, 
which have been cross-checked against the earlier parametrisations of 
Refs.~\cite{Freitas:2013dpa,Freitas:2014hra,Dubovyk:2018rlg}.

The prediction for the $W$-boson mass is obtained by solving the on-shell relation 
that connects $M_W$ to the precisely measured Fermi constant $G_F$, including 
radiative corrections through $\Delta r$. We use the parametrisation of 
Ref.~\cite{Awramik:2003rn}, based on the complete electroweak two-loop 
result~\cite{Freitas:2000gg,Awramik:2003ee,Freitas:2002ja,Awramik:2002wn,
Onishchenko:2002ve} and supplemented by the known three- and four-loop QCD 
corrections~\cite{Schroder:2005db,Chetyrkin:2006bj,Boughezal:2006xk}. This on-shell 
result agrees within about $4\mev$ with the independent two-loop $\overline{\rm MS}$ 
calculation of Ref.~\cite{Degrassi:2014sxa}, which is consistent with the estimated 
missing-higher-order uncertainty of $4\mev$ on $M_W$~\cite{Awramik:2003rn}. The 
recent calculation of the mixed electroweak-QCD corrections of order $\alpha^2\alpha_s$ 
to $\Delta r$~\cite{Dubovyk:2026nhx}, not yet included in the fit, shifts the 
$M_W$ prediction by about $3\mev$, confirming that this uncertainty estimate 
is realistic and not overly conservative.
The $W$-boson width is known at one electroweak loop order, and we use the 
parametrisation of Ref.~\cite{Cho:2011rk}. The corresponding theoretical 
uncertainty remains well below the experimental precision.
\begin{table}[t]
\centering
\begin{tabular}{l@{\hskip 2cm}r}
\hline\noalign{\smallskip}
Quantity & Uncertainty  \\
\noalign{\smallskip}\hline\noalign{\smallskip}
$\deltatheo\MW$                  & $4.0\mev$          \\
$\deltatheo\sinleff$             & $4.3\cdot10^{-5}$  \\
$\deltatheo\sinbeff$             & $5.3\cdot10^{-5}$  \\
$\deltatheo\sinfeff$             & $4.7\cdot10^{-5}$  \\
$\deltatheo R_{\mathrm{NS}}$     & $80.0\,\as^4(M_Z)$     \\
$\deltatheo R_{\mathrm{SA}}$     & $87.6\,\as^4(M_Z)$     \\
\noalign{\smallskip}\hline
\end{tabular}\hspace{1.2cm}
\begin{tabular}{l@{\hskip 2cm}r}
\hline\noalign{\smallskip}
Quantity & Uncertainty\\
\noalign{\smallskip}\hline\noalign{\smallskip}
$\deltatheo\Gamma_{e,\mu,\tau}$  & $0.018\mev$    \\
$\deltatheo\Gamma_\nu$           & $0.016\mev$    \\
$\deltatheo\Gamma_{u,c}$         & $0.11\mev$     \\
$\deltatheo\Gamma_{d,s}$         & $0.08\mev$     \\
$\deltatheo\Gamma_b$             & $0.18\mev$     \\
$\deltatheo\sigma^0_{\rm had}$   & $6\:\mathrm{pb}$ \\
\noalign{\smallskip}\hline
\end{tabular}
\caption{Theoretical uncertainties from missing higher-order corrections included 
as nuisance parameters in the fit. See text for details. \label{tab:theounc}
}
\end{table}

Theoretical uncertainties associated with missing higher-order corrections are 
implemented in Gfitter as Gaussian nuisance parameters. For the $Z$-boson partial 
widths and the hadronic peak cross section, we adopt the estimates from the most 
recent electroweak two-loop calculation~\cite{Dubovyk:2019szj}. As discussed above, 
the prediction of $M_W$ is assigned an uncertainty of $4\mev$. The uncertainties 
on the effective weak mixing angle predictions are $\deltatheo\sinleff = 4.3\cdot10^{-5}$ for 
charged leptons and $\deltatheo\sinbeff = 5.3\cdot10^{-5}$ for $b$ quarks~\cite{Dubovyk:2019szj}, 
while for neutrinos, light quarks and the $c$ quark we use 
$\deltatheo\sinfeff = 4.7\cdot10^{-5}$~\cite{Awramik:2006uz}. The value of $\deltatheo\sinleff$ 
is 9\% smaller than in our previous analysis~\cite{Haller:2018nnx}, owing to the 
updated two-loop calculation of Ref.~\cite{Dubovyk:2019szj}, which includes 
$\Order(\alpha m_t^2 \as^3)$ corrections~\cite{Schroder:2005db,Chetyrkin:2006bj,
Boughezal:2006xk}. The complete set of nuisance parameters describing theoretical 
uncertainties in the fit is summarised in Table~\ref{tab:theounc}.

For quantities derived from partial widths, such as $\Gamma_Z$ and $R^0_f$, the 
theoretical uncertainties are obtained by combining the corresponding $\deltatheo\Gamma_i$. 
This gives $\deltatheo\Gamma_Z = 0.4\mev$ and $\deltatheo R_\ell^0 = 0.006$.

QCD corrections to EWPO enter through the absorptive part of the electromagnetic-current 
correlator, encoded in $R = \sigma(\ee \to \mathrm{hadrons}) / \sigma(\ee \to \mm)$.
They are known up to $\Order(\as^4)$ for the numerically dominant non-singlet (NS) 
contribution~\cite{Baikov:2008jh} and for the smaller singlet contributions from vector 
(SV) and axial-vector (SA) currents~\cite{Baikov:2012er}. Since these $\Order(\as^4)$ 
effects are small compared with the current experimental precision, we take the full 
$\Order(\as^4)$ corrections as theoretical uncertainties. Varying the NS contribution, 
$\deltatheo R_{\mathrm{NS}}$, gives uncertainties of $0.27\mev$ in $\Gamma_Z$ and 
$0.003$ in $R_\ell^0$, and varying the SA contribution gives $0.08\mev$ and $0.001$, 
respectively. The SV contribution has an even smaller numerical impact than the 
$\as^4$ SA term, and no additional $\deltatheo R_{\mathrm{SV}}$ uncertainty is included.


\subsection{Results}
\label{sec:sm}

\begin{table}
\setlength{\tabcolsep}{0.0pc}
{\small
\begin{tabular*}{\textwidth}{@{\extracolsep{\fill}}lccccc} 
\hline\noalign{\smallskip}
& & Free &  & \multicolumn{1}{c}{w/o exp.\ input} & \multicolumn{1}{c}{w/o exp.\ input}   \\[-0.1cm]
\rs{Parameter} & \rs{Input value} & in fit & \rs{Fit Result} & \multicolumn{1}{c}{in line} & \multicolumn{1}{c}{in line, no theo.\ unc} \\
\noalign{\smallskip}\hline\noalign{\smallskip}
$M_{H}$ {\ft [GeV]} &  $125.13\pm0.11$ & yes & $125.13\pm0.11$ & $112^{+19}_{-17}$ & $112^{+16}_{-15}$\\
\noalign{\smallskip}\hline\noalign{\smallskip}
$M_{W}$ {\ft [GeV]} &  $80.3625\pm0.0077$ & -- &  $80.3584\pm0.0048$ &  $80.3558\pm0.0061$ &  $80.3558\pm0.0036$\\
$\Gamma_{W}$ {\ft [GeV]} &  $2.140\pm0.050$ & -- &  $2.090\pm0.001$ &  $2.090\pm0.001$ &  $2.090\pm0.001$\\
\noalign{\smallskip}\hline\noalign{\smallskip}
$M_{Z}$ {\ft [GeV]} &  $91.1879\pm0.0020$ & yes &  $91.1882\pm0.0019$ &  $91.1938\pm0.0069$ &  $91.1937\pm0.0060$\\
$\Gamma_{Z}$ {\ft [GeV]} &  $2.4955\pm0.0023$ & -- &  $2.4945\pm0.0006$ &  $2.4945\pm0.0006$ &  $2.4944\pm0.0005$\\
$\sigma_{\rm had}^{0}$ {\ft [nb]} &  $41.480\pm0.033$ & -- &  $41.490\pm0.008$ &  $41.491\pm0.008$ &  $41.491\pm0.005$\\
$R^{0}_{\l}$ &  $20.767\pm0.025$ & -- &  $20.750\pm0.008$ &  $20.748\pm0.008$ &  $20.748\pm0.005$\\
$A_{\rm FB}^{0,\l}$ &  $0.0171\pm0.0010$ & -- &  $0.01626\pm0.0001$ &  $0.01625\pm0.0001$ &  $0.01626\pm0.0001$\\
$A_\ell$ $^{(\star)}$  & $0.1499\pm0.0018$ & --  & $0.1472\pm0.0004$ & $0.1472\pm0.0004$ & $0.1472\pm0.0003$\\
$\sinleff(Q_{\rm FB})$ &  $0.2324\pm0.0012$ & -- &  $0.23149\pm0.00005$ &  $0.23149\pm0.00005$ &  $0.23149\pm0.00003$\\
$\sinleff(\rm HC)$ &  $0.23149\pm0.00021$ & -- &  $0.23149\pm0.00005$ &  $0.23150\pm0.00005$ &  $0.23150\pm0.00003$\\
$A_{s}$ &  $0.895\pm0.091$ & -- &  $0.9357\pm0.00004$ &  $0.9357\pm0.00004$ &  $0.9357\pm0.00002$\\
$A_{c}$ &  $0.670\pm0.027$ & -- &  $0.6679\pm0.00020$ &  $0.6679\pm0.00020$ &  $0.6679\pm0.00011$\\
$A_{b}$ &  $0.923\pm0.020$ & -- &  $0.93475\pm0.00004$ &  $0.93475\pm0.00004$ &  $0.93475\pm0.00002$\\
$A_{\rm FB}^{0,c}$ &  $0.0707\pm0.0035$ & -- &  $0.0738\pm0.0002$ &  $0.0738\pm0.0002$ &  $0.0738\pm0.0001$\\
$A_{\rm FB}^{0,b}$ &  $0.0996\pm0.0016$ & -- &  $0.1032\pm0.0003$ &  $0.1033\pm0.0003$ &  $0.1033\pm0.0002$\\
$R^{0}_{c}$ &  $0.1721\pm0.0030$ & -- &  $0.17220\pm0.00006$ &  $0.17220\pm0.00006$ &  $0.17220\pm0.00002$\\
$R^{0}_{b}$ &  $0.21629\pm0.00066$ & -- &  $0.21589\pm0.00009$ &  $0.21588\pm0.00009$ &  $0.21588\pm0.00001$\\
\noalign{\smallskip}\hline\noalign{\smallskip}
$\mc$ {\ft [GeV]} &  $1.273\pm0.003$ & yes &  $1.273\pm0.003$ & --  & -- \\
$\mb$ {\ft [GeV]} &  $4.183\pm0.004$ & yes &  $4.183\pm0.004$ & --  & -- \\
$m_{t}$ {\ft [GeV]}$^{(\bigtriangledown)}$ &  $172.52\pm0.60$ & yes &  $172.67\pm0.56$ &  $173.6\pm1.5$ & $173.6\pm1.3$\\
$\dalphaHadMZ$ $^{(\dag\bigtriangleup)}$ &  $2758\pm  10$ & yes & $2757\pm   9$ & $2739\pm  33$ & $2738\pm  30$\\
$\alpha_{s}(M_{Z}^{2})$ &  $0.1177\pm0.0009$ & yes &  $0.1179\pm0.0009$ &  $0.1199\pm0.0028$ &  $0.1198\pm0.0027$\\
\noalign{\smallskip}\hline
\noalign{\smallskip}
\end{tabular*}
\par\vspace{1mm}
\begingroup
\footnotesize
\setlength{\baselineskip}{9.5pt}
\setlength{\parindent}{0pt}
\setlength{\parskip}{0pt}
\noindent
\textsuperscript{(\(\star\))} Average of LEP ($A_\ell=0.1465\pm0.0033$) and SLD
($A_\ell=0.1513\pm0.0021$) measurements, used as two measurements in the fit.
The fit without the LEP measurement gives $A_\ell=0.1472\pm0.0004$, and without
SLD $A_\ell=0.1471\pm0.0004$.
\textsuperscript{(\(\bigtriangledown\))} Combination of experimental ($0.33\gev$)
and theoretical ambiguity uncertainty ($0.5\gev$).
\textsuperscript{(\(\dag\))} In units of $10^{-5}$.
\textsuperscript{(\(\bigtriangleup\))} Rescaled due to the $\alpha_s$ dependence.
\par
\endgroup
}
\caption{
Input values and fit results for the parameters and observables of the global
electroweak fit. The first column lists the fit parameters and observables, and
the second column gives their experimental determinations; the corresponding
references and the treatment of theoretical uncertainties are given in
Section~\ref{sec:theory}. The third column indicates whether a parameter is
free to vary in the fit. The fourth column shows the results of the full fit
using all experimental and theoretical inputs, with uncertainties obtained from
the $\Delta\chi^2=1$ profiles. The fifth column (``w/o exp. input in line'')
gives the indirect determination of each quantity, \ie the result of a fit that
does not use the corresponding input measurement in that row. The last column
shows the same indirect determination, but with all theoretical uncertainties
ignored. Dashes indicate quantities that are not determined indirectly.
  \label{tab:results}
}
\end{table}

The global electroweak fit uses the input observables listed in the first two columns of 
Table~\ref{tab:results}. The free parameters are the Higgs and $Z$-boson masses, the charm, 
bottom, and top-quark masses, the hadronic contribution to the running of the electromagnetic coupling, and the strong coupling constant at the $Z$ pole. These are complemented by the 
nuisance parameters describing the theoretical uncertainties discussed in Section~\ref{sec:theory}. 
The results of the full fit are given in the fourth column of Table~\ref{tab:results}, 
with uncertainties obtained from the $\Delta\chi^2=1$ profiles.

The fit yields $\ChiMin=13.8$ for $\Ndof = 17$ degrees of freedom, corresponding to a 
$p$-value of $0.68$, improving on our previous result of $\ChiMin = 18.6$ for $\Ndof = 15$ 
and $p=0.23$. This improvement is driven by the combined LHC measurements of $M_W$ and 
\sinleff, the updated value of $\sigma_{\rm had}^0$ accounting for changes in the LEP 
luminosity~\cite{Voutsinas:2019hwu} and the improved Bhabha cross section~\cite{Janot:2019oyi}, 
and the completed full two-loop calculations~\cite{Dubovyk:2018rlg,Dubovyk:2019szj,
Bernreuther:2016ccf} of the EWPO. It is notable, given that some key measurements 
have uncertainties reduced by up to a factor of two compared with our previous analysis.

\begin{figure}
{\centering
\includegraphics[width=0.469\textwidth]{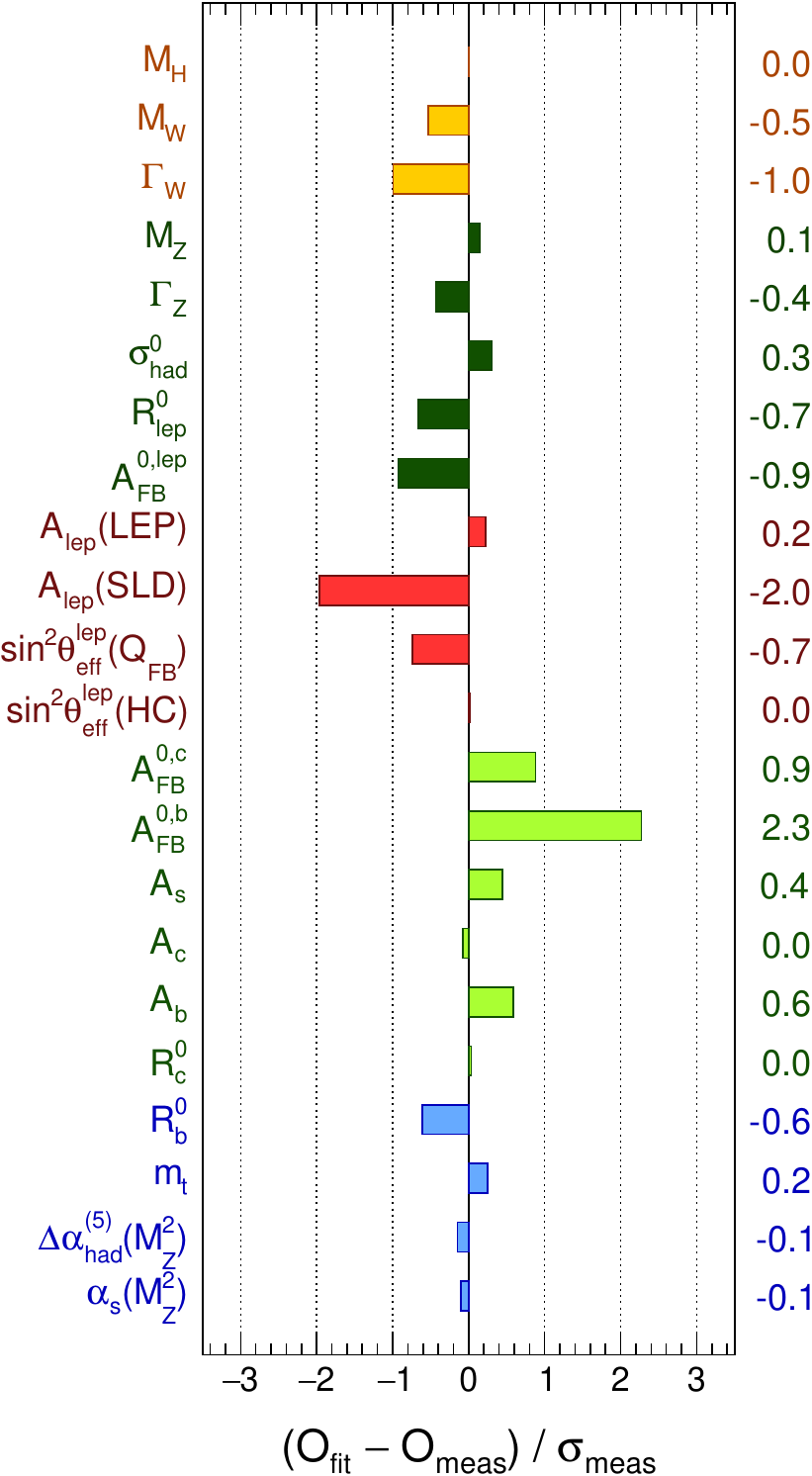}\hspace{0.4cm}
\raisebox{-0.0cm}{\includegraphics[width=0.497\textwidth]{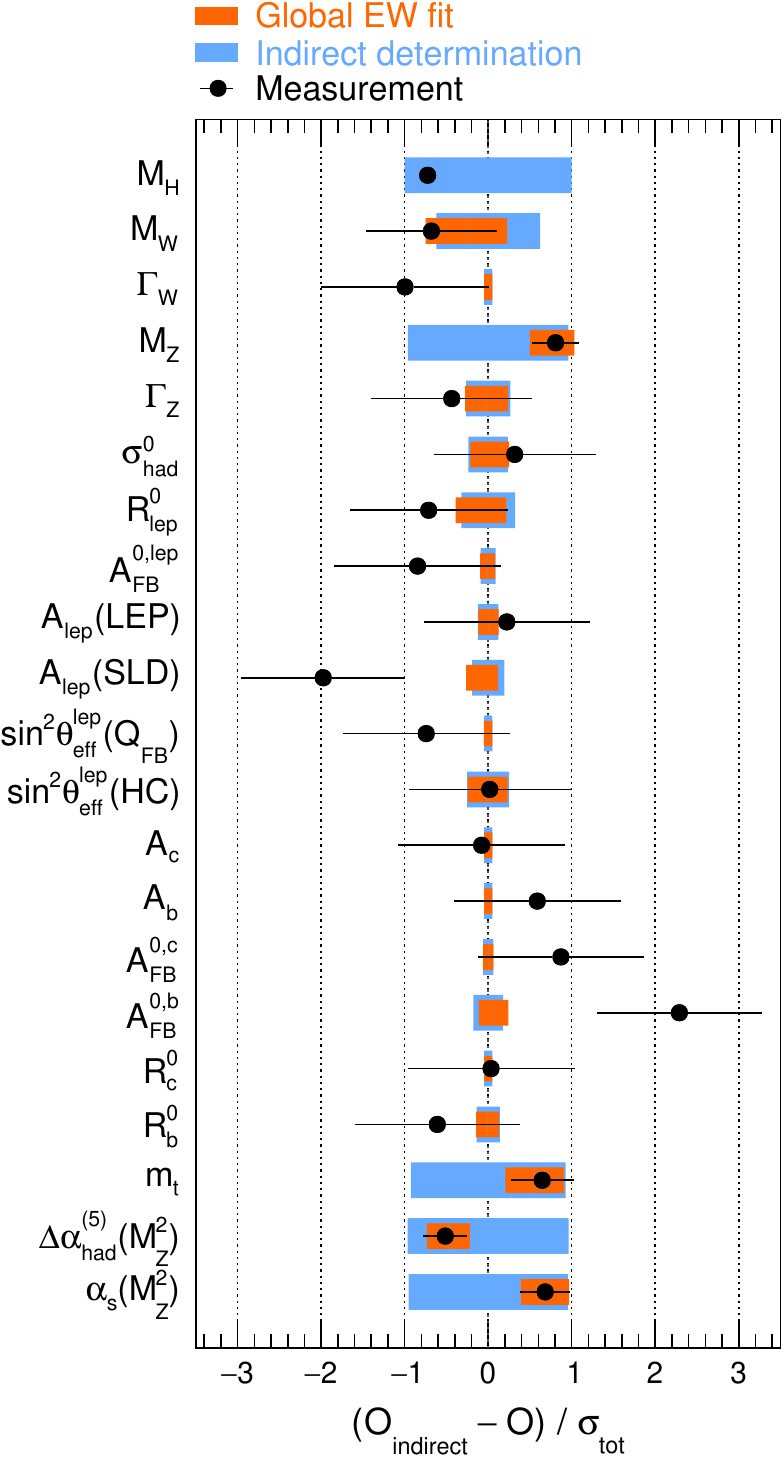}} 
}
\vspace{\captionverticalgap}
\caption[]{Left: pulls in the global electroweak fit, defined as the full-fit result 
minus the input measurement, in units of the measurement uncertainty. Right:
differences between the indirect determinations (blue bands) and the corresponding 
measurements (data points), and between the indirect determinations and the 
full-fit results (orange bands), expressed in units of the quadratic sum of the 
relevant uncertainties.
}
\label{fig:pulls}
\end{figure}

The agreement between the full fit and each input measurement is quantified by the 
pulls shown in Fig.~\ref{fig:pulls} (left). The pulls are defined as the difference 
between the fit result, given in the fourth column of Table~\ref{tab:results}, and 
the input measurement, given in the second column, in units of the measurement 
uncertainty. As in previous analyses, the largest tension is observed in the 
forward–backward asymmetry of $b$ quarks, $A^{0,b}_{\rm FB}$, with a pull of  
$2.3\sigma$, followed by the leptonic asymmetry parameter $A_\ell(\mathrm{SLD})$, 
with a pull of $-2.0\sigma$. All other observables are reproduced within 
about one standard deviation. The pull of $\Gamma_W$ has changed from $0.1\sigma$ 
in our previous fit to $-1.0\sigma$ in the present fit as a consequence of the 
ATLAS measurement, $\Gamma_W=2.202\pm 0.047\gev$~\cite{ATLAS:2024erm}, which 
differs by about $2\sigma$ from the SM prediction, as discussed above.

The right-hand panel of Fig.~\ref{fig:pulls} shows the differences between the
indirect determinations and, respectively, the corresponding measurements
and full-fit results. The indirect determinations are obtained from fits in which
the corresponding measurements are removed. They are listed in the fifth
column of Table~\ref{tab:results} and are in many cases more precise than the
measurements, illustrating the predictive power of the fit.

The indirect determination of the $W$-boson mass reads
\begin{flalign}
  M_W = 80.3558 &\pm 0.0024_{\delta M_Z} \pm 0.0019_{\delta m_t} \pm 0.0029_{\deltaambi{}m_t}
                 \pm 0.0016_{\delta \Delta\alpha_{\rm had}} \nonumber \\[-0.05cm]
                &\pm 0.0008_{\delta \as}  
                 \pm 0.0040_{\deltatheo M_W} \gev\,, \nonumber \\ 
      = 80.3558 &\pm 0.0061_{\mathrm{tot}}\gev\,,
\label{eq:mw_indirect}
\end{flalign}
where the contributions of individual uncertainties show the post-fit {\em impacts}~\cite{Pinto:2023yob}. 
The indirect determination of $M_W$ agrees with the measurement combination, $80.3625\pm0.0077\gev$, within
$0.7\sigma$, and is almost matched in precision. Uncertainties from the input parameters, 
including the top-mass ambiguity uncertainty $\deltaambi{}m_t$, contribute $4.6\mev$, close to the
$4\mev$ theoretical uncertainty. Compared with our previous indirect 
determination, $80.354\pm0.007\gev$~\cite{Haller:2018nnx}, the precision 
improves mainly through the external constraint on $\asZ$ in the fit.

The indirect determination of the effective leptonic weak mixing angle is obtained from a 
fit excluding all measurements with direct sensitivity to it, \ie the left-right asymmetries, 
the forward-backward asymmetries and the direct measurements of \sinleff. 
The result,  
\begin{flalign}
  \sinleff = (23149.4 &\pm 1.1_{\delta M_Z} \pm 0.7_{\delta m_t} \pm 1.1_{\deltaambi m_t} \pm 3.0_{\delta \Delta\alpha_{\rm had}} \nonumber \\[-0.05cm]
                      &\pm 0.6_{\delta \as}  
                       \pm 4.3_{\deltatheo \sinleff} ) \cdot 10^{-5}\,, \nonumber \\
           = 0.231494 &\pm 0.000056_{\mathrm{tot}}\,,
\label{eq:sineff_indirect}
\end{flalign}
agrees with the direct measurements and is considerably more precise. The uncertainty 
is dominated by missing higher-order corrections, followed by \dalphaHadMZ.
Compared with our previous result, $0.23153\pm0.00006$~\cite{Haller:2018nnx},
the precision improves slightly due to the external constraint on \asZ and the
smaller theoretical uncertainty, $\deltatheo \sinleff = 4.3\cdot 10^{-5}$ instead
of $4.7\cdot 10^{-5}$.

The top-quark mass is indirectly determined to be
\beq
  \mt = 173.6 \pm 1.5\gev\,,
\label{eq:mt}
\eeq
in agreement with the direct measurement, $172.52\pm0.60\gev$, within $0.6\sigma$. 
The precision is considerably improved compared to our previous result, $176.4\pm2.1\gev$, 
mainly due to the more precise input on $M_W$. The theoretical uncertainties contribute 
about $0.6\gev$ to the total uncertainty. With perfect knowledge of $M_W$, the total 
uncertainty would be reduced to $0.84\gev$.

\begin{figure}[tb]
{\centering
\includegraphics[width=\defaultSingleFigureScale\textwidth]{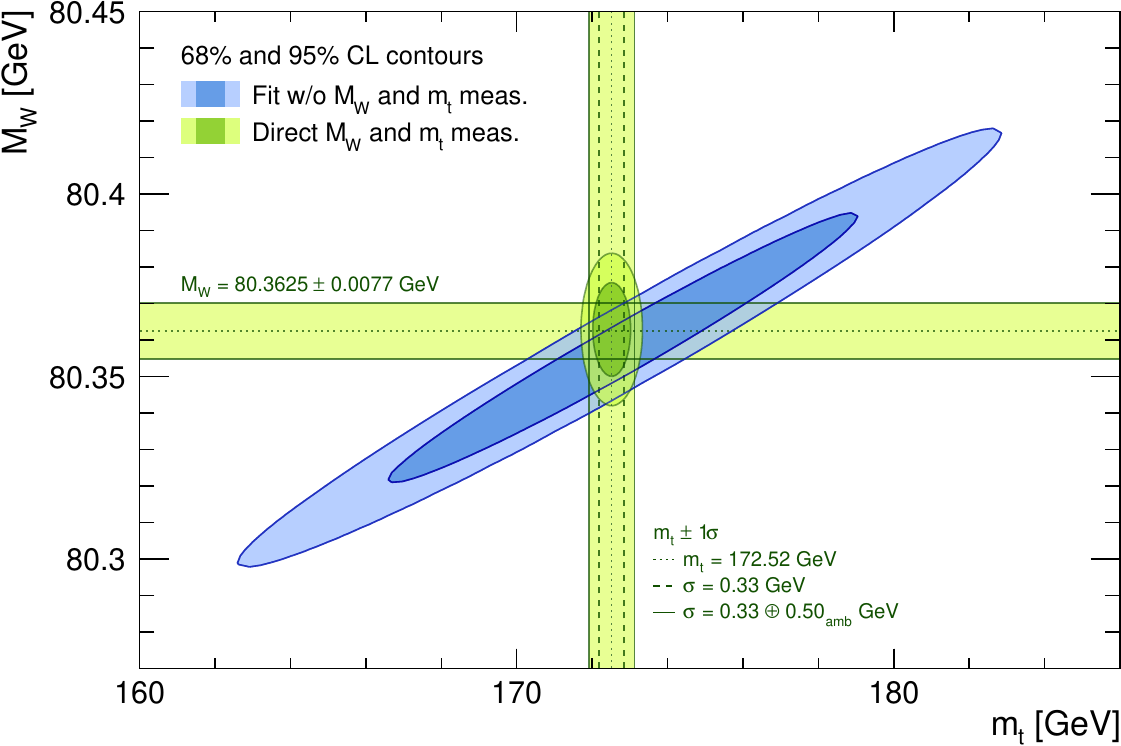} \\
}
\vspace{\captionverticalgap}
\caption[]{Indirect fit determination of $m_t$ and $M_W$(blue), compared with the
experimental values (green).
The 68\% and 95\% confidence-level contours are shown.
\label{fig:MWvsmt}}
\end{figure}
A stringent consistency test of the SM is provided by the simultaneous indirect
determination of $m_t$ and $M_W$. Figure~\ref{fig:MWvsmt} shows the two-dimensional
scan in the $m_t$--$M_W$ plane.
The indirect determination is in remarkable agreement with the direct 
measurements, and both have substantially improved in precision compared with our 
previous results~\cite{Haller:2018nnx}.

The Higgs-boson mass is indirectly determined to be
\beq
  M_H = 112\,^{+19}_{-17}\gev\,,
\label{eq:mh}
\eeq
in agreement with the measured value, $125.1 \pm 0.1\gev$. The slight tension
observed in our previous analysis is reduced from $1.7\sigma$ to $0.7\sigma$
mainly owing to the LHC measurements of $M_W$ and \sinleff.
Using only the ATLAS $M_W$ measurement~\cite{ATLAS:2024erm} gives $M_H = 104^{+36}_{-29}\gev$, 
while the CMS $M_W$ measurement~\cite{CMS:2024lrd} gives $M_H = 116^{+27}_{-23}\gev$. 
The hadron-collider combination \sinleff{}(HC) gives $M_H = 121^{+69}_{-48}\gev$.

\begin{figure}[tb]
\centering
\includegraphics[width=\defaultSingleFigureScale\textwidth]{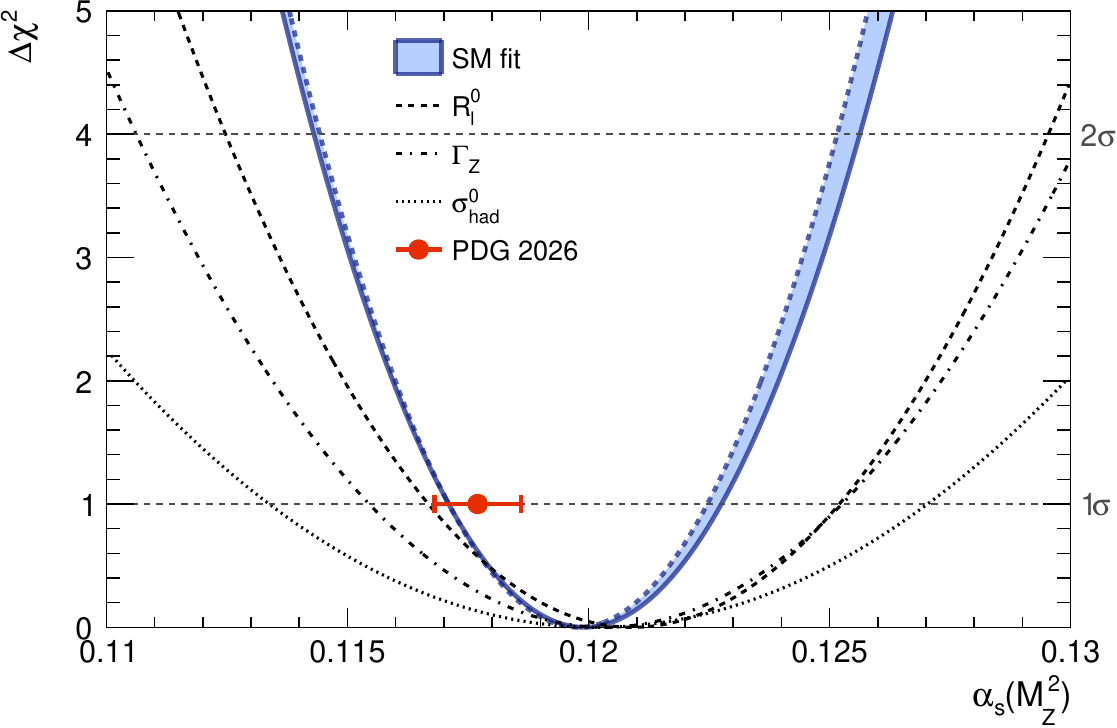}
\vspace{\captionverticalgap}
\caption[]{Scans of \DeltaChi as a function of \asZ, obtained without the external
\asZ constraint. The blue band shows the full-fit result, with its width
indicating the contribution from theoretical uncertainties; solid and dotted
lines correspond to fits with and without these uncertainties. Grey lines show
the determinations from $R^{0}_{\ell}$, $\Gamma_Z$, and $\sigma_{\rm had}^{0}$
alone. The external input value used in the nominal fit, Eq.~\eqref{eq:alphas_pdg},
is shown for comparison.\label{fig:alphas_scan}}
\end{figure}
The fit also determines the strong coupling constant at the $Z$-mass scale
with full electroweak next-to-next-to-leading order (NNLO) and $\Order(\alpha_s^4)$ 
QCD accuracy~\cite{Kniehl:1988ie,
Kniehl:1989qu,Chetyrkin:1993yp,Larin:1993ju,Chetyrkin:1993ug,Chetyrkin:1994js, Baikov:2008jh,Baikov:2012er} to be 
\beq
\asZ = 0.1199 \pm 0.0028\,.
\label{eq:alphas}
\eeq
Theoretical uncertainties have little impact on this determination; omitting them gives
$\asZ = 0.1198 \pm 0.0027$. The most sensitive observables are $R^{0}_{\ell}$,
$\Gamma_Z$, and $\sigma_{\rm had}^{0}$, giving individual uncertainties of
$0.0042$, $0.0049$, and $0.0068$, respectively. Figure~\ref{fig:alphas_scan} 
shows the \DeltaChi profiles for these observables, which yield mutually consistent 
\asZ determinations agreeing within 0.0007. The updated $\sigma_{\rm had}^{0}$ value, 
incorporating the improved Bhabha cross section~\cite{Janot:2019oyi} and revised LEP 
luminosity treatment~\cite{Voutsinas:2019hwu}, shifts the \asZ determination from 
this observable upward by about $1\sigma$, with only a small impact on the combination.

Overall, the updated global electroweak fit shows excellent consistency, with no new
significant discrepancies beyond the known tensions in $A_\ell(\mathrm{SLD})$ and
$A_{\rm FB}^{0,b}$. The improved direct measurements and indirect determinations
test the SM with unprecedented precision.

\subsection{Oblique parameters}

\begin{figure}
\centering
\includegraphics[width=\defaultSingleFigureScale\textwidth]{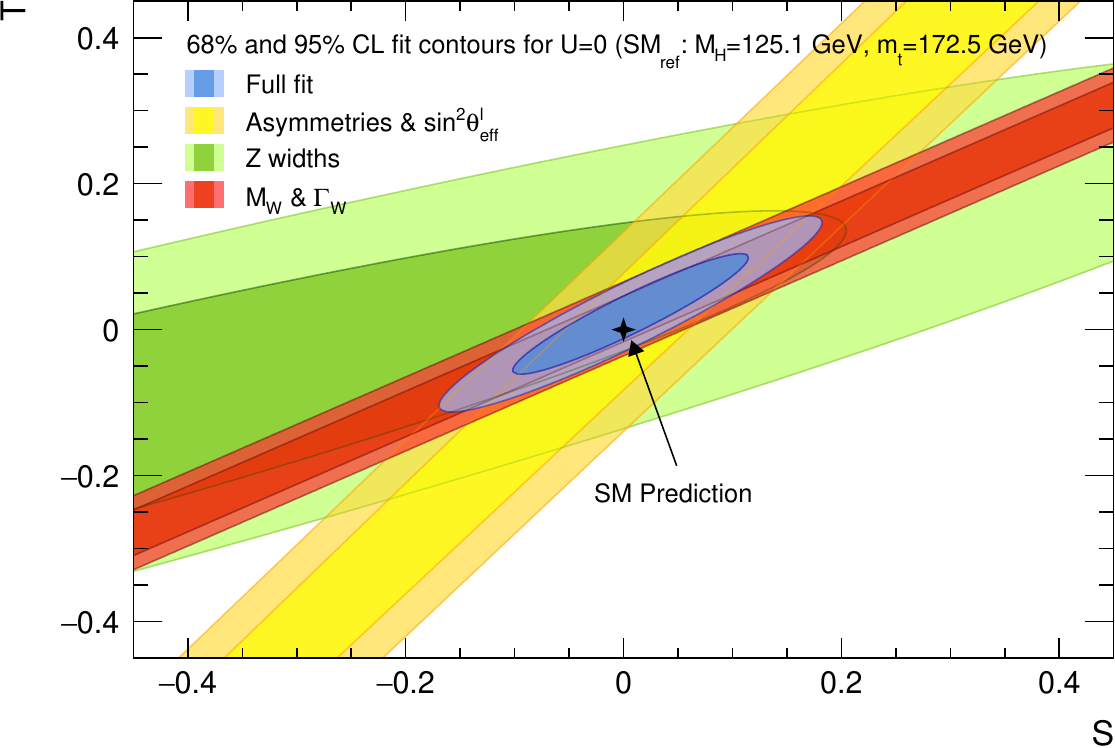} 
\vspace{\captionverticalgap}
\caption[]{Constraints on the oblique parameters $S$ and $T$, with $U$ fixed to zero.
The combined constraint from all observables is shown in blue. Individual constraints 
are shown for the asymmetry and direct \sinleff measurements (yellow), the $Z$ partial 
and total widths (green), and the $W$ mass and width (red), with confidence levels 
drawn for one degree of freedom.\label{fig:STU}}
\end{figure}
Using the SM reference values $M_{H,{\rm ref}}=125\gev$ and $m_{t,{\rm ref}}=172.5\gev$, 
we determine the oblique parameters $S$, $T$, and $U$, which parametrise possible new-physics contributions to electroweak gauge-boson self-energies~\cite{Peskin:1990zt,
Peskin:1991sw}.\footnote{The parameter $S$ mainly captures new physics that modifies 
the momentum dependence of neutral-current self-energies; $T$ encodes custodial-symmetry 
breaking through differences between neutral and charged-current self-energies at zero 
momentum; and $U$ quantifies differences in the momentum dependence of charged and 
neutral-current self-energies.} 
We find
\beq
  S= \SParam\,, \hspace{0.5cm}
  T= \TParam\,, \hspace{0.5cm}
  U= \UParam\,,
\label{eq:stu}
\eeq 
with correlation coefficients $\STParamCor$ between $S$ and $T$,
$\SUParamCor$ between $S$ and $U$, and $\TUParamCor$ between $T$ and $U$.
When fixing $U=0$, the fit gives $S|_{U=0}= \SParamNU$ and $T|_{U=0}= \TParamNU$, with 
a correlation of $\STParamCorNo$. The corresponding constraints in the $S$--$T$ 
plane are shown in Fig.~\ref{fig:STU}. The LHC measurements of $M_W$ strongly constrain 
one linear combination of $S$ and $T$; complementary precision observables are needed
to close the allowed region and provide both upper and lower bounds. Owing to the new 
measurements of $M_W$ and \sinleff, the central values are closer to the SM expectation 
and the uncertainties are reduced with respect to our previous analysis~\cite{Haller:2018nnx}.

\section{The Higgs sector}
\label{sec:higgs}

The Higgs-boson coupling measurements at the LHC provide a complementary set of
precision tests of the SM. While EWPO constrain the Higgs sector through its virtual 
contributions to gauge-boson self-energies, direct measurements of Higgs-boson 
production and decay rates probe the same sector through observable rates. In this 
section, we combine these two sources of information in the $\kappa$ 
framework~\cite{LHCHiggsCrossSectionWorkingGroup:2013rie}: the constraints on the 
Higgs coupling to electroweak gauge bosons obtained from the electroweak fit,
encoded through the oblique parameters, are combined with the ATLAS and CMS
Higgs signal-strength measurements. This improves the determination of Higgs
couplings and provides additional sensitivity to the total Higgs-boson width and
to possible invisible or undetected decay modes.

\subsection{Signal strength measurements}

We use Higgs-boson signal strength modifiers, $\mu$, to constrain the Higgs-boson 
couplings. For each production mode and decay channel, $\mu$ is defined as the ratio of 
the observed yield to the SM prediction. The inputs are the most recent ATLAS~\cite{ATLAS:2025qxq}
and CMS~\cite{CMS:2026nce} signal-strength combinations, including their
correlations within each experiment. Correlations between ATLAS and CMS 
are neglected. The measurements entering these combinations are summarised below.

In the $H\to\gamma\gamma$ channel, which benefits from a clean signature despite
its small branching fraction, ATLAS provides separate signal strengths for
\ttH, \ggH, \VBF, \WH, and \ZH production~\cite{ATLAS:2022tnm}. We do not use the
single-top associated production mode, \tH, since it introduces interference
terms not included in the parametrisations below and its impact is negligible. 
CMS measured the same decay channel for \ggH, \VBF, \WH, \ZH, and combined \ttHtH
production~\cite{CMS:2021kom}. Since the \tH contribution is small, this
combined channel is treated as \ttH throughout.

The clean $H\to ZZ$ channel, with both $Z$ bosons decaying leptonically ($e$, $\mu$), is 
measured by ATLAS for \ggH, \VBF, \VH, and \ttHtH production~\cite{ATLAS:2020rej}, with
additional \ttHtH information from a dedicated multilepton analysis~\cite{ATLAS:2017ztq}.
CMS provides signal strength measurements for \ggH, \VBF, \WH, \ZH, and \ttHtH 
production~\cite{CMS:2021ugl}, complemented by a dedicated \ttHtH analysis~\cite{CMS:2020mpn}.

In the $H\to WW$ channel, with both $W$ bosons decaying leptonically, ATLAS
measured \ggH and \VBF~\cite{ATLAS:2025hki}, as well as \WH and
\ZH~\cite{ATLAS:2025abg} production, complemented by \ttHtH results from
Ref.~\cite{ATLAS:2017ztq}. CMS measured \ggH, \VBF, \WH, and \ZH
production~\cite{CMS:2022uhn}, and released a dedicated \ttHtH
search in Ref.~\cite{CMS:2020mpn}.

For $H\to\tau\tau$, ATLAS measured signal strengths for \ggH, \VBF, and 
\ttHtH~\cite{ATLAS:2024wfv} as well as \WH and \ZH~\cite{ATLAS:2023qpu} production,
with additional \ttHtH information from Ref.~\cite{ATLAS:2017ztq}. CMS measured
\ggH, \VBF, \WH, and \ZH production~\cite{CMS:2022kdi}, supplemented by a
dedicated \ttHtH search~\cite{CMS:2020mpn}.

For $H\to\bbbar$, the dominant decay mode, we include ATLAS measurements in 
\WH, \ZH~\cite{ATLAS:2024yzu}, \VBF~\cite{ATLAS:2020bhl}, 
and \ttHtH production~\cite{ATLAS:2024gth}. CMS inputs include \WH and
\ZH~\cite{CMS:2023vzh}, \ttHtH~\cite{CMS:2024fdo}, \ggH, and \VBF 
production~\cite{CMS:2024ddc}, and an additional VBF-only
analysis~\cite{CMS:2023tfj}.

For $H\to\ccbar$, we include the ATLAS \WH and \ZH results~\cite{ATLAS:2024yzu}. 
Since this channel is not included in the CMS combination~\cite{CMS:2026nce}, we 
add the CMS \ttH~\cite{CMS:2025dsh} and combined \VH results~\cite{CMS:2022psv}, 
assuming no correlation with the other signal strength measurements.

For $H\to\mu\mu$, we use ATLAS~\cite{ATLAS:2024wfv} and CMS~\cite{CMS:2020xwi}
measurements combining \ggH, \ttHtH, \VBF, and \VH production. For
$H\to Z\gamma$, we include the inclusive ATLAS measurement~\cite{ATLAS:2020qcv}
and the CMS measurements in \ggH and \VBF production~\cite{CMS:2022ahq}.

All signal-strength uncertainties are symmetrised by averaging the upper and
lower uncertainties. If a result is bounded at $\mu=0$, the lower uncertainty is
set equal to the upper one to avoid underestimating the total uncertainty. This
allows negative $\mu$ values within confidence intervals, reflecting the large
uncertainties of low-sensitivity channels in a frequentist treatment. Choosing 
instead the larger of the upper and lower uncertainties for symmetrisation
does not significantly alter the results.

\subsection{Theoretical framework}
\label{sec:kappasandmeas}

To parametrise the Higgs-boson couplings to SM particles, we use coupling
modifiers \kappai~\cite{LHCHiggsCrossSectionWorkingGroup:2013rie}. For a
production mode $i$ and decay channel $f$, the signal strength is written as
\beq
  \mu_{i}^{f} = \frac{\kappai^2 \cdot \kappaf^2}{\kappaH^2}\,.
  \label{eq:higgsSignalStrength}
\eeq
Here, $\kappai^2$ and $\kappaf^2$ scale the production cross section in mode $i$
and the partial decay width into the final state $f$, respectively, while
$\kappaH^2$ denotes the total-width modifier,
\beq
  \kappa^2_H = \frac{\GammaHtot}{\GammaHSM}\,.
  \label{eq:kappaH2}
\eeq
In addition to the known SM decay modes, the total width may receive contributions
from invisible or undetected final states,
\beq
  \GammaBSM = \Gamma_{\mathrm{inv}} + \Gamma_{\mathrm{undet}}\,,
  \label{eq:Hiu}
\eeq
where invisible final states do not interact with the detector, as in the case of
neutrinos or dark-matter particles, while undetected final states escape experimental
identification, for example because of inefficient reconstruction or large
irreducible backgrounds. The total width is then given by
\begin{equation}
  \label{eq:GammaHtot-GammaBSM}
  \GammaHtot = \GammaH + \GammaBSM \,,
\end{equation}
with
\beq
  \GammaH = \sum_i \Gamma_i(\setkappa)
          = \sum_i \kappa_i^2 \, \Gamma^{\mathrm{SM}}_i\,,
  \label{eq:gammaH}
\eeq
for a given set of coupling modifiers $\setkappa$. The SM prediction,
corresponding to $\GammaBSM = 0$ and $\kappai = 1$, is
$\GammaHSM = 4.10\mev$~\cite{LHCHiggsCrossSectionWorkingGroup:2016ypw}.

We consider several parametrisations with different sets of \kappai modifiers,
corresponding to different assumptions. This allows us to present both
model-independent results and results with stronger constraints under more
restrictive assumptions. Depending on the parametrisation, different subsets of
the available measurements are included in the fit. The parametrisations
considered are summarised in Table~\ref{tab:kappa}.
\afterpage{\begin{landscape} 
\begin{table}[p] 
\setlength{\tabcolsep}{0.5pc} 
\begin{center}
{\scriptsize
\begin{tabular}{@{\extracolsep{\fill}}llcccc} 
\hline\noalign{\smallskip} 
Production  &  Decay  &  Bosonic  &  Effective  &  Resolved  &  General  \\ 
\noalign{\smallskip}\hline\noalign{\smallskip} 
\multirow{7}{*}{\ggH}  &  $\gamma\gamma$  &  ---  &  $(0.99\kappag^2+0.01) \kappaga^2/\kappaH^2$  &  $\kappaF^2 (0.068\kappaF^2-0.657\kappaF\kappaV+1.589\kappaV^2)/\kappaH^2$  &  $(0.99\kappag^2+0.01\kappab^2) \kappaga^2/\kappaH^2$  \\ 
 &  $ZZ$  &  ---  &  $(0.99\kappag^2+0.01) \kappaV^2/\kappaH^2$  &  $\kappaF^2\kappaV^2/\kappaH^2$  &  $(0.99\kappag^2+0.01\kappab^2) \kappaV^2/\kappaH^2$  \\ 
 &  $WW$  &  ---  &  $(0.99\kappag^2+0.01) \kappaV^2/\kappaH^2$  &  $\kappaF^2\kappaV^2/\kappaH^2$  &  $(0.99\kappag^2+0.01\kappab^2) \kappaV^2/\kappaH^2$  \\ 
 &  $\tau\tau$  &  ---  &  $(0.99\kappag^2+0.01) \kappatau^2/\kappaH^2$  &  $\kappaF^4/\kappaH^2$  &  $(0.99\kappag^2+0.01\kappab^2) \kappatau^2/\kappaH^2$  \\ 
 &  $\bbbar$  &  ---  &  ---  &  $\kappaF^2\kappaV^2/\kappaH^2$  &  $(0.99\kappag^2+0.01\kappab^2) \kappab^2/\kappaH^2$  \\ 
 &  $\mu\mu$  &  ---  &  ---  &  $\kappaF^4/\kappaH^2$  &  $(0.99\kappag^2+0.01\kappab^2) \kappamu^2/\kappaH^2$  \\ 
 &  $Z\gamma$  &  ---  &  ---  &  $\kappaF^2 (0.003\kappaF^2-0.121\kappaF\kappaV+1.118\kappaV^2)/\kappaH^2$  &   $(0.99\kappag^2+0.01\kappab^2) \kappaZga^2/\kappaH^2$  \\ 
\noalign{\smallskip}\hline\noalign{\smallskip} 
\multirow{7}{*}{VBF}  &  $\gamma\gamma$  &  ---  &  $\kappaV^2 \kappaga^2/\kappaH^2$  &  $\kappaV^2 (0.068\kappaF^2-0.657\kappaF\kappaV+1.589\kappaV^2)/\kappaH^2$  &  $\kappaV^2 \kappaga^2/\kappaH^2$  \\ 
 &  $ZZ$  &  $\kappaV^4/\kappaH^2$  &  $\kappaV^4/\kappaH^2$  &  $\kappaV^4/\kappaH^2$  &  $\kappaV^4/\kappaH^2$  \\ 
 &  $WW$  &  $\kappaV^4/\kappaH^2$  &  $\kappaV^4/\kappaH^2$  &  $\kappaV^4/\kappaH^2$  &  $\kappaV^4/\kappaH^2$  \\ 
 &  $\tau\tau$  &  ---  &  $\kappaV^2 \kappatau^2/\kappaH^2$  &  $\kappaV^2\kappaF^2/\kappaH^2$  &  $\kappaV^2 \kappatau^2/\kappaH^2$  \\ 
 &  $\bbbar$  &  ---  &  ---  &  $\kappaV^2\kappaF^2/\kappaH^2$  &  $\kappaV^2 \kappab^2/\kappaH^2$  \\ 
 &  $\mu\mu$  &  ---  &  ---  &  $\kappaV^2\kappaF^2/\kappaH^2$  &  $\kappaV^2 \kappamu^2/\kappaH^2$  \\ 
 &  $Z\gamma$  &  ---  &  ---  &  $\kappaV^2 (0.003\kappaF^2-0.121\kappaF\kappaV+1.118\kappaV^2)/\kappaH^2$  &  $\kappaV^2 \kappaZga^2/\kappaH^2$  \\ 
\noalign{\smallskip}\hline\noalign{\smallskip} 
\multirow{7}{*}{$WH$/$ZH$/$VH$}  &  $\gamma\gamma$  &  ---  &  $\kappaV^2 \kappaga^2/\kappaH^2$  &  $\kappaV^2 (0.068\kappaF^2-0.657\kappaF\kappaV+1.589\kappaV^2)/\kappaH^2$  &  $\kappaV^2 \kappaga^2/\kappaH^2$  \\ 
 &  $ZZ$  &  $\kappaV^4/\kappaH^2$  &  $\kappaV^4/\kappaH^2$  &  $\kappaV^4/\kappaH^2$  &  $\kappaV^4/\kappaH^2$  \\ 
 &  $WW$  &  $\kappaV^4/\kappaH^2$  &  $\kappaV^4/\kappaH^2$  &  $\kappaV^4/\kappaH^2$  &  $\kappaV^4/\kappaH^2$  \\ 
 &  $\tau\tau$  &  ---  &  $\kappaV^2 \kappatau^2/\kappaH^2$  &  $\kappaV^2\kappaF^2/\kappaH^2$  &  $\kappaV^2 \kappatau^2/\kappaH^2$  \\ 
 &  $\bbbar$  &  ---  &  ---  &  $\kappaV^2\kappaF^2/\kappaH^2$  &  $\kappaV^2 \kappab^2/\kappaH^2$  \\ 
 &  $\ccbar$  &  ---  &  ---  &  $\kappaV^2\kappaF^2/\kappaH^2$  &  $\kappaV^2 \kappac^2/\kappaH^2$  \\ 
 &  $\mu\mu$  &  ---  &  ---  &  $\kappaV^2\kappaF^2/\kappaH^2$  &  $\kappaV^2 \kappamu^2/\kappaH^2$  \\ 
\noalign{\smallskip}\hline\noalign{\smallskip} 
\multirow{7}{*}{$\ttbar H(+tH)$}  &  $\gamma\gamma$  &  ---  &  ---  &  $\kappaF^2 (0.068\kappaF^2-0.657\kappaF\kappaV+1.589\kappaV^2)/\kappaH^2$  &  $\kappat^2 \kappaga^2/\kappaH^2$  \\ 
 &  $ZZ$  &  ---  &  ---  &  $\kappaF^2\kappaV^2/\kappaH^2$  &  $\kappat^2 \kappaV^2/\kappaH^2$  \\ 
 &  $WW$  &  ---  &  ---  &  $\kappaF^2\kappaV^2/\kappaH^2$  &  $\kappat^2 \kappaV^2/\kappaH^2$  \\ 
 &  $\tau\tau$  &  ---  &  ---  &  $\kappaF^4/\kappaH^2$  &  $\kappat^2 \kappatau^2/\kappaH^2$  \\ 
 &  $\bbbar$  &  ---  &  ---  &  $\kappaF^4/\kappaH^2$  &  $\kappat^2 \kappab^2/\kappaH^2$  \\ 
 &  $\ccbar$  &  ---  &  ---  &  $\kappaF^4/\kappaH^2$  &  $\kappat^2 \kappac^2/\kappaH^2$  \\ 
 &  $\mu\mu$  &  ---  &  ---  &  $\kappaF^4/\kappaH^2$  &  $\kappat^2 \kappamu^2/\kappaH^2$  \\ 
\noalign{\smallskip}\hline\noalign{\smallskip} 
$ggH$+VBF  &  $\bbbar$  &  ---  &  ---  &  $(0.928\kappaF^2+0.072\kappaV^2)\kappaF^2/\kappaH^2$  &  $(0.9187\kappag^2+0.0093\kappab^2+ 0.072\kappaV^2 ) \kappab^2/\kappaH^2$  \\ 
VBF+$VH$  &  $\mu\mu$  &  ---  &  ---  &  $\kappaV^2\kappaF^2/\kappaH^2$  &  $\kappaV^2 \kappamu^2/\kappaH^2$  \\ 
$ggH$+$\ttbar H$+$tH$  &  $\mu\mu$  &  ---  &  ---  &  $\kappaF^4/\kappaH^2$  &  $(0.9781\kappag^2 + 0.0099\kappab^2+0.0120\kappat^2)\kappamu^2/\kappaH^2$  \\[0.1cm] 
\multirow{2}{*}{Inclusive}  &  \multirow{2}{*}{$Z\gamma$}  &  \multirow{2}{*}{---}  &  \multirow{2}{*}{---}  &  $(0.89\kappaF^2+0.11\kappaV^2)(0.003\kappaF^2$ & $(0.8712\kappag^2+0.0088\kappab^2+0.01\kappat^2$ \\[-0.05cm]
 & & & & $-0.121\kappaF\kappaV+1.118\kappaV^2)/\kappaH^2$  &  $+0.11\kappaV^2)\kappaZga^2/\kappaH^2$  \\ 
\noalign{\smallskip}\hline 
\noalign{\smallskip} 
\end{tabular} 
} 
\end{center}
\caption{Parameterisations of the signal strengths in terms of the \kappai couplings as a function of the production and decay modes. 
\label{tab:kappa}} 
\end{table} 
\end{landscape} 
}

{\bf Bosonic parametrisation}
This parametrisation allows only the \HVV coupling to
vary, with $\kappaZ=\kappaW=\kappaV$. This avoids assumptions on fermion
couplings and loop-induced contributions. Accordingly, we use only \WH, \ZH,
\VH, and \VBF production, and only the $H\to WW$ and $H\to ZZ$ decay channels.
This choice restricts the analysis to tree-level modifications of Higgs-boson
production and decay.

{\bf Effective parametrisation.}
Loop-induced production and decay modes are included by introducing \kappag and
\kappaga as free parameters. Among the fermion couplings, only the coupling to
the $\tau$ lepton is considered, parametrised by \kappatau. In addition to the
measurements used in the bosonic parametrisation, we include \ggH production and 
the $H\to\gamma\gamma$ and $H\to\tau\tau$ decay channels. This incorporates some 
of the most precise signal-strength measurements without imposing further assumptions 
on the Higgs coupling structure. The loop-induced contributions are tested for 
deviations from the SM predictions through \kappag and \kappaga, while simultaneously 
constraining the \HVV coupling and $\kappa_H^2$.

{\bf Resolved parametrisation.}
We use a resolved parametrisation with two $\kappa$ modifiers to include all
signal-strength measurements: \kappaV modifies all bosonic couplings and \kappaF 
all fermionic couplings. The \ggH production mode is scaled by \kappaF, assuming 
that only SM fermions contribute to the \ggH loop, while the loop-induced 
$H\to\gamma\gamma$ and $H\to Z\gamma$ decays are modified by the corresponding 
SM combinations of \kappaV and \kappaF~\cite{LHCHiggsCrossSectionWorkingGroup:2013rie}. 
This parametrisation uses the full set of measurements, but imposes the strong 
assumption that all bosonic and all fermionic couplings scale universally.

{\bf General parametrisation.}
This parametrisation introduces the largest set of \kappai modifiers.
All fermion couplings and loop-induced contributions are assigned independent
modifiers, without assumptions on relations between them. As in the effective
parametrisation, the \ggH and $H\to\gamma\gamma$ loop contributions vary
through \kappag and \kappaga, with an additional modifier $\kappaZga$
for the $H\to Z\gamma$ decay. Only the electroweak gauge-boson couplings are
described by a common modifier, \kappaV, to allow the inclusion of EWPO in the
fit, as described below. This parametrisation uses the full set of signal-strength
measurements and constrains the Higgs-boson couplings to all SM particles
simultaneously.

In all parametrisations except the bosonic one, some signal-strength measurements
combine production modes that do not map onto one \kappai modifier. In these
cases, we use cross-section-weighted combinations of the relevant \kappai
modifiers, with weights from the theoretical 
predictions~\cite{LHCHiggsCrossSectionWorkingGroup:2016ypw}.
The numerical combinations are given in Table~\ref{tab:kappa}. In the sum of
partial widths entering $\Gamma_H$, all \kappai modifiers not included in a given
parametrisation are fixed to their SM values, $\kappai=1$.

\subsection{Higgs boson coupling modifiers}
\label{sec:higgscouplings}

\begin{figure}[t!]
\centering
\includegraphics[width=0.6\textwidth]{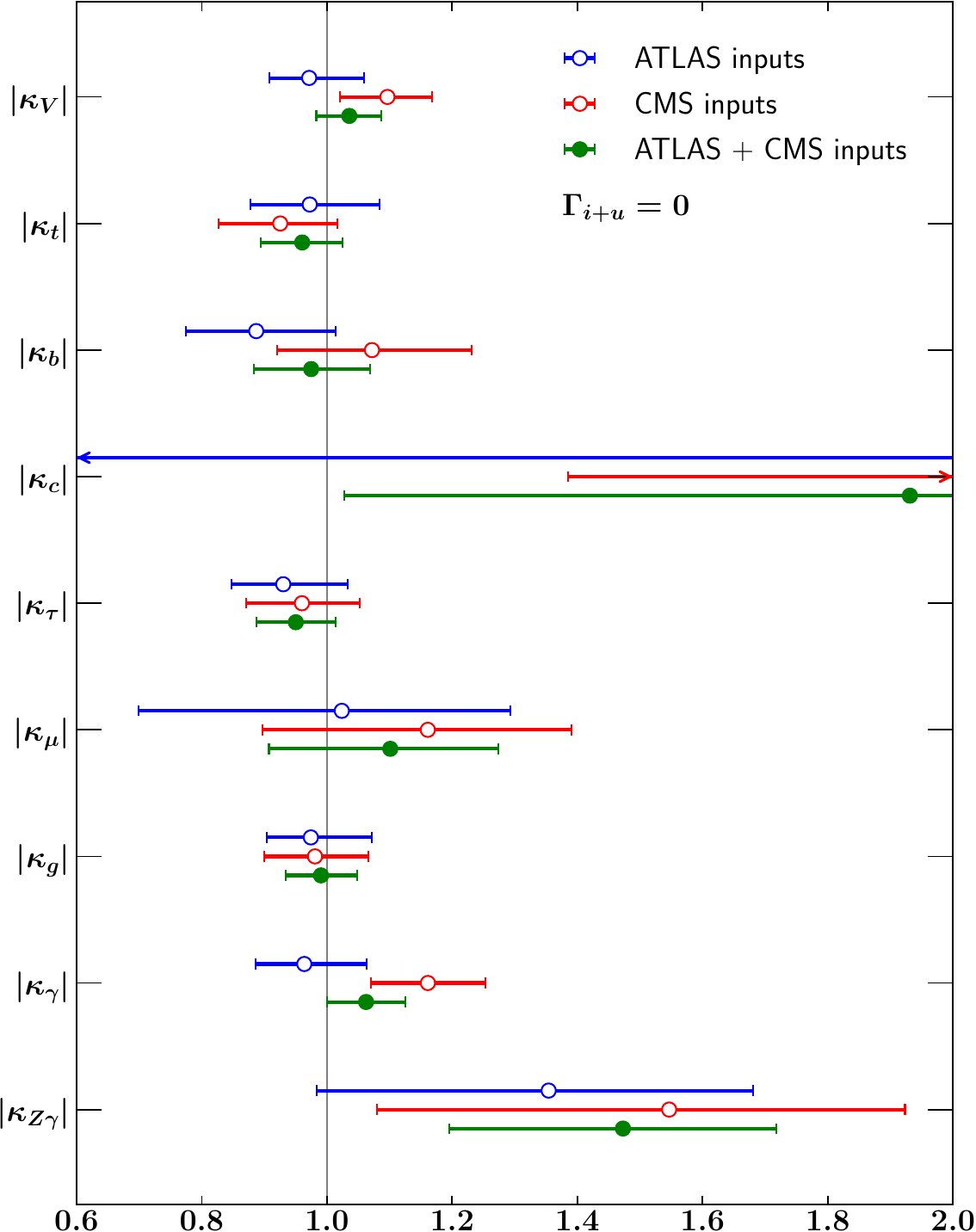} 
\vspace{0.4cm}
\caption[]{Constraints on the \kappai modifiers in the {\rm general parametrisation} 
with $\GammaBSM = 0$, using Higgs signal-strength measurements from ATLAS (blue), 
CMS (red), and their combination (green). Since all \kappai enter quadratically, 
only their absolute values are constrained. Arrows indicate central values outside
the plotting range.
\label{fig:allkappas}}
\end{figure}
We first use the {\em general parametrisation} with the full set of signal-strength
measurements, computing $\Gamma_H$ according to Eq.~\eqref{eq:gammaH}. This
validates our signal-strength implementation and determines the \kappai modifiers
from Higgs data alone, before adding EWPO constraints. We reproduce the published
ATLAS~\cite{ATLAS:2025qxq} and CMS~\cite{CMS:2026nce} \kappai results to very
good accuracy. The ATLAS-only and CMS-only fits, together with their combination,
are shown in Fig.~\ref{fig:allkappas}. All results are mostly consistent with
the SM expectation, $\kappai=1$. For the \HVV modifier, the combined fit gives
$\kappaV = 1.035\,^{+0.051}_{-0.053}$. The largest deviation is found for
$\kappaZga = 1.47\,^{+0.25}_{-0.28}$, corresponding to about $1.7\sigma$.

\begin{figure}[tbp]
\centering
\includegraphics[width=\defaultSingleFigureScale\textwidth]{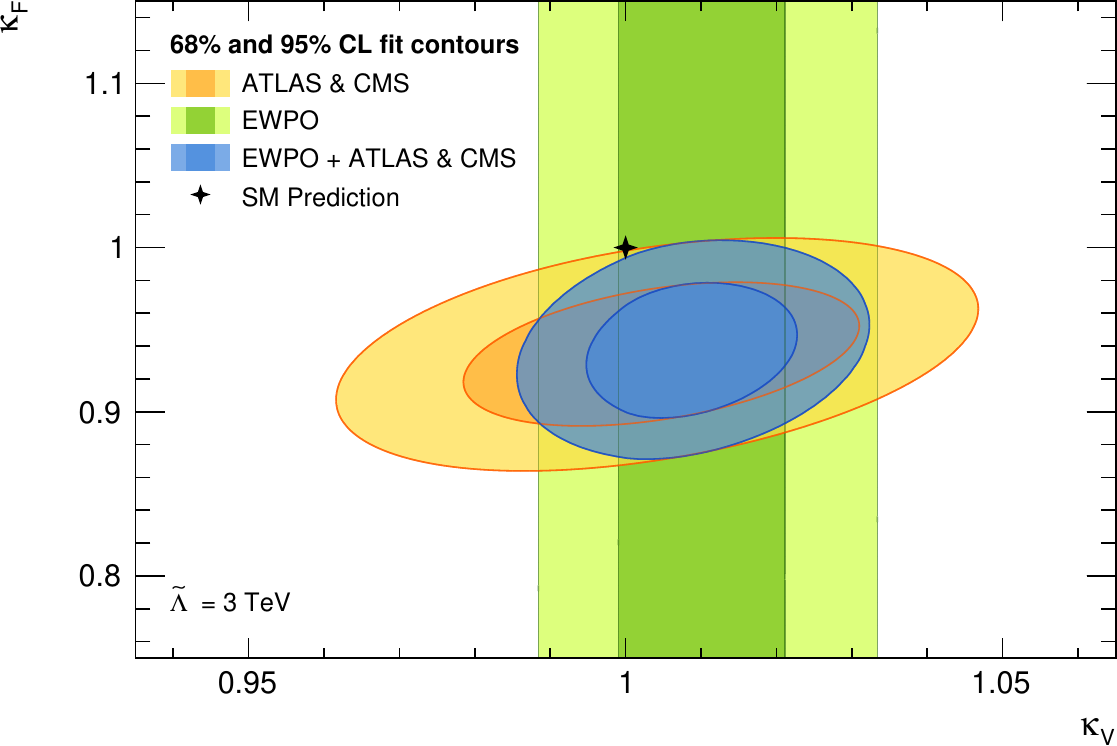} 
\vspace{\captionverticalgap}
\caption[]{Constraints in the $\kappaV$--$\kappaF$ plane from Higgs signal-strength
measurements in the {\em resolved parametrisation} with $\GammaBSM = 0$ (orange), 
EWPO (green), and their combination (blue). In fits including EWPO, the cutoff scale 
is set to $\widetilde{\Lambda}=3\tev$. The 68\% and 95\% contours are computed for 
one degree of freedom in the EWPO-only fit, and for two degrees of freedom otherwise.
\label{fig:kappaVkappaF}}
\end{figure}

We next adopt the {\em resolved parametrisation}, which constrains \kappaV and
\kappaF simultaneously from the full signal-strength set, as shown by the orange
contours in Fig.~\ref{fig:kappaVkappaF} for $\GammaBSM=0$. This result is
combined with the independent \kappaV constraint from electroweak precision data, 
following the strategy of Ref.~\cite{Espinosa:2012im}. At leading-logarithmic accuracy, deviations of the
Higgs-boson coupling to vector bosons modify the oblique parameters $S$ and
$T$ as~\cite{Haber:1993wf,Haber:1999zh,He:2001tp}
\beq
  S= \frac{1}{12 \pi} \left(1 - \kappaV^2\right)\ln\!\frac{\Lambda_{\star}^2}{\MH^2}\,, \hspace{0.9cm}
  T= -\frac{3}{16 \pi c_W^2} \left(1 - \kappaV^2\right)\ln\!\frac{\Lambda_{\star}^2}{\MH^2}\,,
\label{eq:stu_kappaV}
\eeq
with $U=0$. Here, $\Lambda_{\star}=\widetilde{\Lambda}/|1-\kappaV^2|^{1/2}$ is the 
cutoff scale of the effective description, which decreases as the deviation from the 
SM value increases. As default, we use $\widetilde{\Lambda}=4\pi v\simeq3\tev$, 
corresponding to the naive-dimensional-analysis estimate of the strong-coupling scale 
of the electroweak symmetry-breaking sector, where $v$ is the vacuum expectation value.
We also quote results for $\widetilde{\Lambda}=1\tev$, close to the partial-wave
unitarity estimate for the breakdown scale of longitudinal vector-boson
scattering when the SM cancellation is modified. Using the EWPO of 
Table~\ref{tab:results}, without Higgs signal-strength inputs, we obtain
\beq
  \kappaV = \kappaVEWPO\,,
  \label{eq:kappaVEWPO}
\eeq
for $\widetilde{\Lambda}=3\tev$, shown by the green band in
Fig.~\ref{fig:kappaVkappaF}. It agrees with the Higgs signal-strength determination, 
$\kappaV=1.005 \pm 0.017$, from the combined input of ATLAS and CMS measurements in the
{\em resolved parameterisation}, 
but is more precise owing to the high experimental and theoretical precision of the 
EWPO and does not need the $\GammaBSM=0$ constraint. 
For $\widetilde{\Lambda}=1\tev$, we find $\kappaV = 1.011\,^{+0.017}_{-0.013}$. 
The combination of the EWPO constraint with the direct Higgs signal-strength measurements 
is shown by the blue contour in Fig.~\ref{fig:kappaVkappaF}.

\subsection{Higgs boson total width}
\label{sec:higgswidth}

\begin{figure}
\centering
\includegraphics[width=\defaultSingleFigureScale\textwidth]{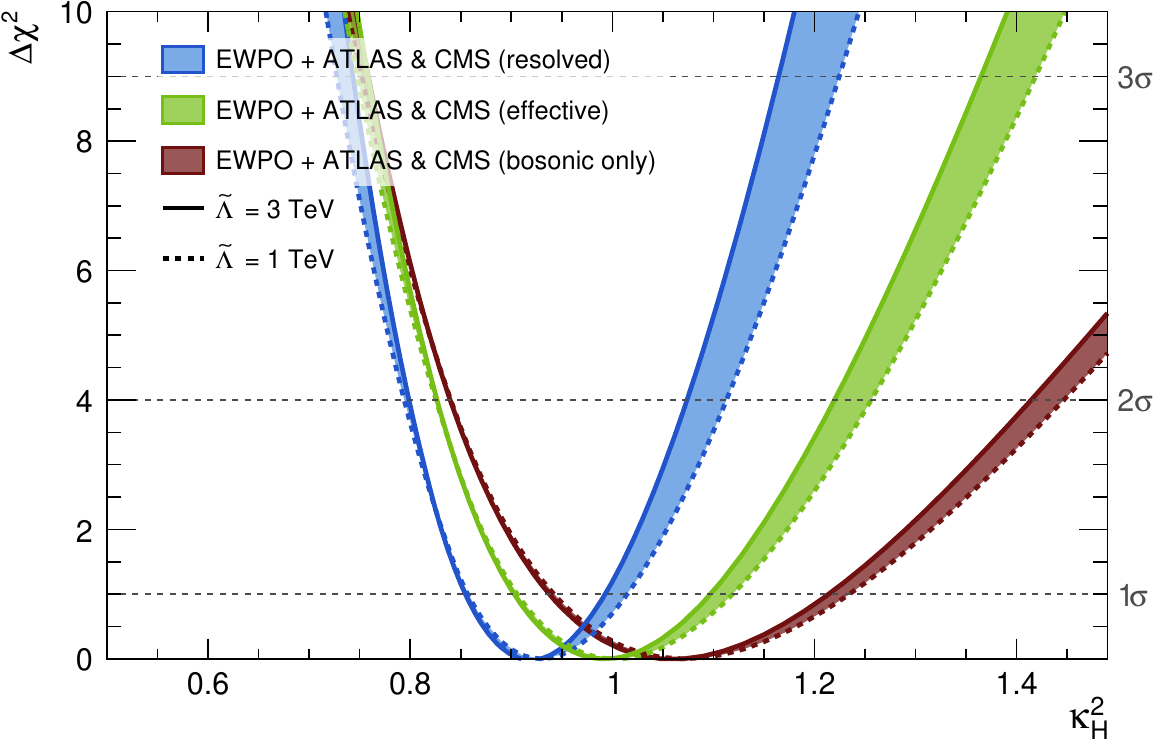} 
\vspace{\captionverticalgap}
\caption[]{Scans of \DeltaChi versus the total Higgs-width modifier $\kappa_H^2$ 
for different parametrisations. Solid and dashed lines correspond to 
$\widetilde{\Lambda}=3\tev$ and $1\tev$, respectively.\label{fig:kappaH2}}
\end{figure}
At the LHC, the Higgs-boson width has been determined indirectly from the ratio
of off-shell to on-shell production rates, under the assumption of SM couplings,
with the results 
$\GammaHtot = 4.3\,^{+2.7}_{-1.9}\mev$ from ATLAS~\cite{ATLAS:2024jry} and
$\GammaHtot = 3.0\,^{+2.0}_{-1.5}\mev$ from CMS~\cite{CMS:2024eka}.
Both are compatible with the SM prediction,
$\GammaHSM = 4.10\pm0.06\mev$~\cite{LHCHiggsCrossSectionWorkingGroup:2016ypw}.
At the current precision, these determinations change only mildly when the
SM-coupling assumption is relaxed in realistic SM extensions~\cite{Stylianou:2026yxn}.
This may change at future Higgs factories~\cite{LinearColliderVision:2025hlt,deBlas:2944678},
whereas at the HL-LHC the limited precision of the total-width determination can
weaken the sensitivity to BSM scenarios~\cite{Forslund:2023reu}.

By providing an independent constraint on \kappaV, the EWPO reduce the 
freedom in the Higgs-coupling fit and enable a determination of the total
Higgs-boson width without specifying a particular SM extension. We treat $\kappaH^2$ 
as a free parameter, rather than imposing Eq.~\eqref{eq:kappaH2}, and determine
$\GammaHtot = \kappaH^2 \GammaHSM$. The resulting constraints on $\kappaH^2$
for different parametrisations and two choices of $\widetilde{\Lambda}$ are shown in
Fig.~\ref{fig:kappaH2}. The numerical results below assume $\widetilde{\Lambda}=3\tev$.

The {\em bosonic parametrisation} includes only tree-level modifications of the \HVV
couplings, using \VBF and \VH production and the $H\to WW$ and $H\to ZZ$ decay
channels. The fit uses 11 signal-strength measurements and the 24 EWPO from
Table~\ref{tab:results}.
In addition to the electroweak-fit parameters, \kappaV and $\kappaH^2$ are left
free. We obtain 
\beq
\GammaHtot = \GammaHHVV,
\eeq
with $\ChiMin/\Ndof = 29.2/26$, corresponding to a $p$-value of 0.3.

The {\em effective parametrisation} includes loop-induced Higgs-boson production and
decay modes, allowing possible BSM contributions through \kappag and \kappaga.
It incorporates additional signal-strength constraints, in particular
the precise $H\to\tau^+\tau^-$ and $H\to\gamma\gamma$ measurements in \ggH
production, without further assumptions. This fit uses 31 signal-strength
measurements to constrain \kappaV, \kappag, \kappaga, \kappatau, and
$\kappaH^2$, together with the EWPO. We obtain
\beq
  \GammaHtot = \GammaHEff\,,
\eeq
with $\ChiMin/\Ndof = 46.3/43$, corresponding to a $p$-value of 0.34.

In the {\em resolved parametrisation}, all 59 signal-strength measurements are 
used to determine \kappaV, \kappaF, and $\kappaH^2$. We find
\beq
\GammaHtot = \GammaHRes\,,
\eeq
with similarly good fit quality, $\ChiMin/\Ndof = 76.9/73$. 

Including more Higgs signal-strength information therefore tightens the 
constraints on the total width, at the cost of stronger assumptions on the 
Higgs-boson couplings. In particular, the {\em effective} and
{\em resolved parametrisations} determine the total width with a relative precision of
about $10\%$ or better. To our knowledge, this is the first determination of
$\GammaHtot$ at this level of precision without relying on strong assumptions
such as $|\kappaV|<1$. 

The results depend only mildly on the cutoff scale $\widetilde{\Lambda}$. Reducing
$\widetilde{\Lambda}$ from $3\tev$ to $1\tev$ shifts the central values by
$-0.04\mev$, independently of the parametrisation. The uncertainties on
$\GammaHtot$ increase by about 5\%, 10\%, and 15\% for the bosonic, effective,
and resolved parametrisations, respectively (\cf size of bands in 
Fig.~\ref{fig:kappaH2}). Adding the ATLAS and CMS off-shell width constraints 
in $H\to ZZ^{*}\to 4\ell$~\cite{ATLAS:2023dnm,CMS:2024eka} does not noticeably 
improve the uncertainty on \GammaHtot, since these constraints are less precise 
than the combined fit.

\subsection{Invisible and undetected width}
\label{sec:higgsinvisible}

Following Eqs.~\eqref{eq:GammaHtot-GammaBSM} and \eqref{eq:gammaH}, Higgs boson 
decays to final states not experimentally accessible at the LHC can be accommodated 
in the total Higgs-boson width via~\cite{LHCHiggsCrossSectionWorkingGroup:2013rie}
\begin{equation}
\GammaHtot = \frac{\GammaH(\setkappa)}{1-\BRHinv-\BRHundet}\,,
\end{equation}
where $\BRHinv=\Gamma_{\mathrm{inv}}/\GammaHtot$ and 
$\BRHundet=\Gamma_{\mathrm{undet}}/\GammaHtot$ denote the invisible and undetected
branching fractions, respectively. The visible width $\GammaH(\setkappa)$ is obtained 
from the SM partial widths scaled by the coupling modifiers. The total-width 
modifier~\eqref{eq:kappaH2} can be rewritten in terms of $\BRBSM=\BRHinv+\BRHundet$ as
\begin{equation}
\kappaH^2 = \frac{\GammaHtot}{\GammaHSM} 
          = \frac{\GammaH(\setkappa)}{\GammaHSM}\frac{1}{1-\BRBSM}\,,
\end{equation}
translating the total-width constraint obtained from the combined EWPO and
Higgs signal-strength fit into a constraint on $\BRBSM$. In this formulation, 
$\BRBSM$ is directly determined from the data and does not rely on the SM prediction 
for the total Higgs-boson width.

\begin{figure}[tb]
\centering
\includegraphics[width=\defaultSingleFigureScale\textwidth]{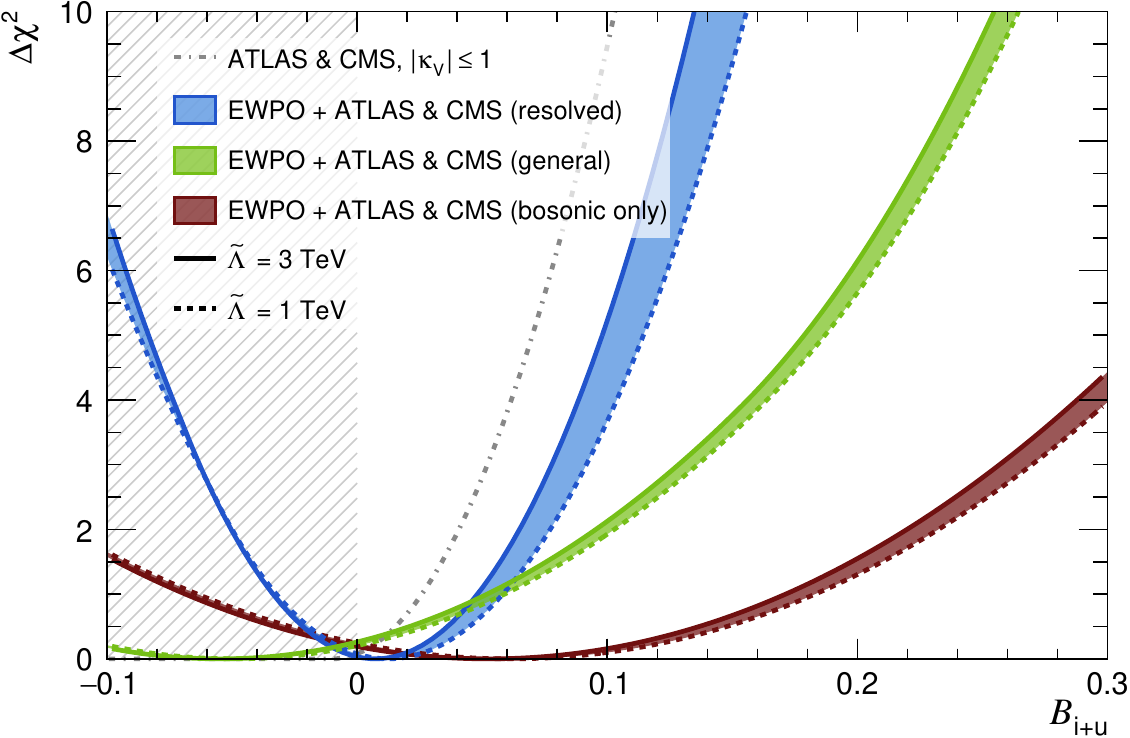} 
\vspace{0.2cm}
\caption[]{Scans of \DeltaChi versus $\BRBSM=\BRHinv+\BRHundet$ from the combined 
fit to EWPO and Higgs signal strengths, shown for different parametrisations. 
Solid and dashed lines correspond to $\widetilde{\Lambda}=3\tev$ and $1\tev$,
respectively. The signal-strength-only result with $|\kappaV|<1$ is shown for
comparison.\label{fig:BRHinv}}
\end{figure}

We fit $\BRBSM$ in the parametrisations defined above, with the results shown in
Fig.~\ref{fig:BRHinv}. The numerical limits quoted below are obtained at 95\% CL
by integrating the likelihood over the physical region $\BRBSM>0$, and using
$\widetilde{\Lambda}=3\tev$. In the {\em bosonic parametrisation}, where only the
\HVV coupling is modified, we find $\BRBSM < \BRBSMVV$ with an expected upper limit 
of 0.23.\footnote{
  The expected limits are calculated under the assumption of the SM, where $\kappaV = 1$ and $\mu_i^f = 1$ 
  with unchanged uncertainties, for all measurements entering the fit. 
}
We note that in this parameterisation, 
all other coupling modifiers, most notably the fermionic ones, are set to their SM values
in the determination of \BRBSM. 
In the {\em general parametrisation}, $\BRBSM$ is fitted together with all \kappai modifiers, giving
$\BRBSM < \BRBSMgen$ with an expected limit of 0.16. 
In this parameterisation, no assumption on any of the coupling modifiers is needed. 
The {\em resolved parametrisation} gives the tighter bound,
$\BRBSM < \BRBSMres$ (0.08 expected), owing to its stronger assumptions on 
fermion couplings and loop-induced processes.

These bounds do not rely on the assumption $|\kappaV|<1$, often used when only
signal-strength measurements are available. Imposing this assumption with signal
strengths alone gives the strongest bound $\BRBSM < \BRBSMnoEWPO$ at 95\% CL.

\subsection{Discussion}


The constraints on \kappaV, \GammaHtot, and \BRBSM derived above rely on a
restricted effective interpretation. The EWPO enter only through the leading
logarithmic contribution of a universal \HVV rescaling to the oblique
parameters $S$ and $T$, with $U=0$. This assumes an otherwise unchanged SM loop
structure, unbroken custodial symmetry, and no BSM effects beyond oblique
corrections. Modified fermion couplings, vertex corrections, and direct loop
contributions from new states to the EWPO are not included.

Since a rescaled \HVV coupling alone is not a complete renormalisable theory, the
cutoff dependence in Eq.~\eqref{eq:stu_kappaV} should be regarded as a
leading-logarithmic estimate. A systematic effective field theory treatment 
requires the relevant operators, counterterms and short-distance assumptions. 
Beyond the one-loop approximation, additional cutoff-sensitive contributions 
may arise unless they are suppressed or cancelled by the UV completion.
The resulting bounds therefore apply to this oblique leading-log scenario, not
to arbitrary BSM physics.

\section{Constraining the SMEFT}
\label{sec:smeft}

\subsection{Effective field theory framework}
\label{sec:smeftframework}

A systematic parametrisation of BSM effects is provided by the Standard Model
Effective Field Theory (SMEFT)~\cite{Brivio:2017vri}, which assumes that possible 
new states are heavier than the electroweak scale and that the Higgs field transforms 
as part of a linearly realised $\rm SU(2)_L$ doublet. The low-energy Lagrangian can then 
be written as
\beq
  \mathcal{L}_{\mathrm{SMEFT}} = \mathcal{L}_{\mathrm{SM}}
  + \sum_i \frac{c_i}{\Lambda^2}\, \mathcal{O}_i^{(6)} + \dots\,,
  \label{eq:smeft_lagrangian}
\eeq
where $\Lambda$ is the characteristic BSM scale, $\mathcal{O}_i^{(6)}$ are
dimension-six operators built from SM fields and invariant under the SM gauge
symmetry, and $c_i$ are dimensionless Wilson coefficients. The ellipsis denotes 
terms of order $\Lambda^{-4}$ and beyond, corresponding to operators of dimension 
eight and higher. The dimension-five operator, which violates lepton number, is not 
considered here. We use the Warsaw basis~\cite{Buchmuller:1985jz,Grzadkowski:2010es}, 
assuming baryon-number conservation and flavour universality, and follow the 
conventions and broken-phase Feynman rules of Ref.~\cite{Dedes:2017zog}.

In the fit, the dimension-six operators modify the two-loop SM predictions of
the EWPO. We use the NLO QCD and electroweak results of
Ref.~\cite{Dawson:2019clf} for the $Z$-pole and $W$ observables in the Warsaw
basis, computed in the on-shell scheme with $\alpha$, $M_Z$, and $G_F$ as inputs.
For an observable ${\cal X}$, the prediction is written as
\beq
  {\cal X} = {\cal X}^{\mathrm{SM}} + \sum_i \frac{c_i}{\Lambda^2}\, \delta {\cal X}_i\,,
  \label{eq:smeft_shift}
\eeq
with the coefficients $\delta {\cal X}_i$ evaluated to NLO. We retain only terms linear
in the Wilson coefficients, corresponding to $\Order(\Lambda^{-2})$, and neglect
quadratic dimension-six terms of $\Order(\Lambda^{-4})$, which are of the same
formal order as dimension-eight interference effects. The shifts in the relations
between input parameters and observables are included following
Ref.~\cite{Dawson:2019clf}, allowing the EWPO of Section~\ref{sec:ewfit} to be
reinterpreted as constraints on $c_i/\Lambda^2$.

\subsection{Bounds on Wilson coefficients}
\label{sec:smeftresults}

\begin{table}[t]
\centering
\small
\setlength{\tabcolsep}{0.5pc}
\begin{tabular}[t]{lcc}
\hline\noalign{\smallskip}
\multirow{2}{*}{$c_i/\Lambda^2$}   & \multicolumn{2}{c}{95\% CL interval $[\tev^{-2}]$} \\
                                   & Full fit &  SM fixed \\
\noalign{\smallskip}\hline\noalign{\smallskip}
$c_{HWB}$ & $[-0.0046, 0.0028]$ & $[-0.0047, 0.0017]$\\
$c_{Hl}^{(3)}$ & $[-0.0073, 0.0038]$ & $[-0.0076, 0.0022]$\\
$c_{Hl}^{(1)}$ & $[-0.0054, 0.0067]$ & $[-0.0053, 0.0060]$\\
$c_{He}$ & $[-0.0084, 0.0081]$ & $[-0.0073, 0.0082]$\\
$c_{HD}$ & $[-0.012, 0.0066]$ & $[-0.012, 0.0038]$\\
$c_{ll}$ & $[-0.0071, 0.013]$ & $[-0.0045, 0.014]$\\
$c_{Hq}^{(3)}$ & $[-0.0087, 0.014]$ & $[-0.0067, 0.014]$\\
$c_{Hq}^{(1)}$ & $[-0.025, 0.039]$ & $[-0.023, 0.037]$\\
$c_{Hu}$ & $[-0.038, 0.072]$ & $[-0.024, 0.075]$\\
$c_{Hd}$ & $[-0.13, 0.053]$ & $[-0.13, 0.040]$\\
$c_{lu}$ & $[-0.28, 0.22]$ & $[-0.25, 0.22]$\\
$c_{uB}$ & $[-0.33, 0.19]$ & $[-0.34, 0.12]$\\
$c_{qq}^{(3)}$ & $[-0.34, 0.24]$ & $[-0.34, 0.19]$\\
$c_{lq}^{(1)}$ & $[-0.30, 0.37]$ & $[-0.29, 0.34]$\\
$c_{lq}^{(3)}$ & $[-0.29, 0.44]$ & $[-0.19, 0.47]$\\
$c_{eu}$ & $[-0.36, 0.37]$ & $[-0.37, 0.33]$\\
\noalign{\smallskip}\hline
\end{tabular}
\hspace{2.5em}
\begin{tabular}[t]{lcc}
\hline\noalign{\smallskip}
\multirow{2}{*}{$c_i/\Lambda^2$}   & \multicolumn{2}{c}{95\% CL interval $[\tev^{-2}]$} \\
                                   & Full fit &  SM fixed \\
\noalign{\smallskip}\hline\noalign{\smallskip}
$c_{uW}$ & $[-0.56, 0.34]$ & $[-0.58, 0.25]$\\
$c_{qe}$ & $[-0.48, 0.47]$ & $[-0.42, 0.48]$\\
$c_{uu}$ & $[-1.2, 0.84]$ & $[-1.2, 0.68]$\\
$c_{qq}^{(1)}$ & $[-1.0, 1.4]$ & $[-1.0, 1.4]$\\
$c_{qu}^{(1)}$ & $[-2.2, 1.3]$ & $[-2.3, 1.1]$\\
$c_{W}$ & $[-2.5, 1.4]$ & $[-2.5, 0.77]$\\
$c_{ee}$ & $[-2.5, 2.4]$ & $[-2.2, 2.5]$\\
$c_{ud}^{(1)}$ & $[-2.5, 5.8]$ & $[-1.9, 5.7]$\\
$c_{ld}$ & $[-4.2, 5.0]$ & $[-4.2, 4.4]$\\
$c_{le}$ & $[-5.4, 7.0]$ & $[-5.1, 6.7]$\\
$c_{qd}^{(1)}$ & $[-10, 4.0]$ & $[-9.9, 3.1]$\\
$c_{ed}$ & $[-7.6, 6.8]$ & $[-6.7, 7.0]$\\
$c_{HW}$ & $[-9.5, 5.1]$ & $[-9.5, 3.0]$\\
$c_{HB}$ & $[-11, 6.3]$ & $[-11, 3.9]$\\
$c_{H\Box}$ & $[-11, 6.1]$ & $[-11, 3.5]$\\
$c_{dd}$ & $[-52, 21]$ & $[-51, 17]$\\
\noalign{\smallskip}\hline
\end{tabular}
\caption{Summary of the 95\% CL intervals on $C_i/\Lambda^2$ from the SMEFT fit,
obtained by varying one operator at a time. The full fit, profiling the SM
inputs and theory uncertainties together with the Wilson coefficients, is given 
in the second column of each table, whereas the fixed-SM configuration is given 
in the third column. \label{tab:smeft}}
\end{table}

\begin{figure}[ptb]
\centering
\includegraphics[width=0.94\textwidth]{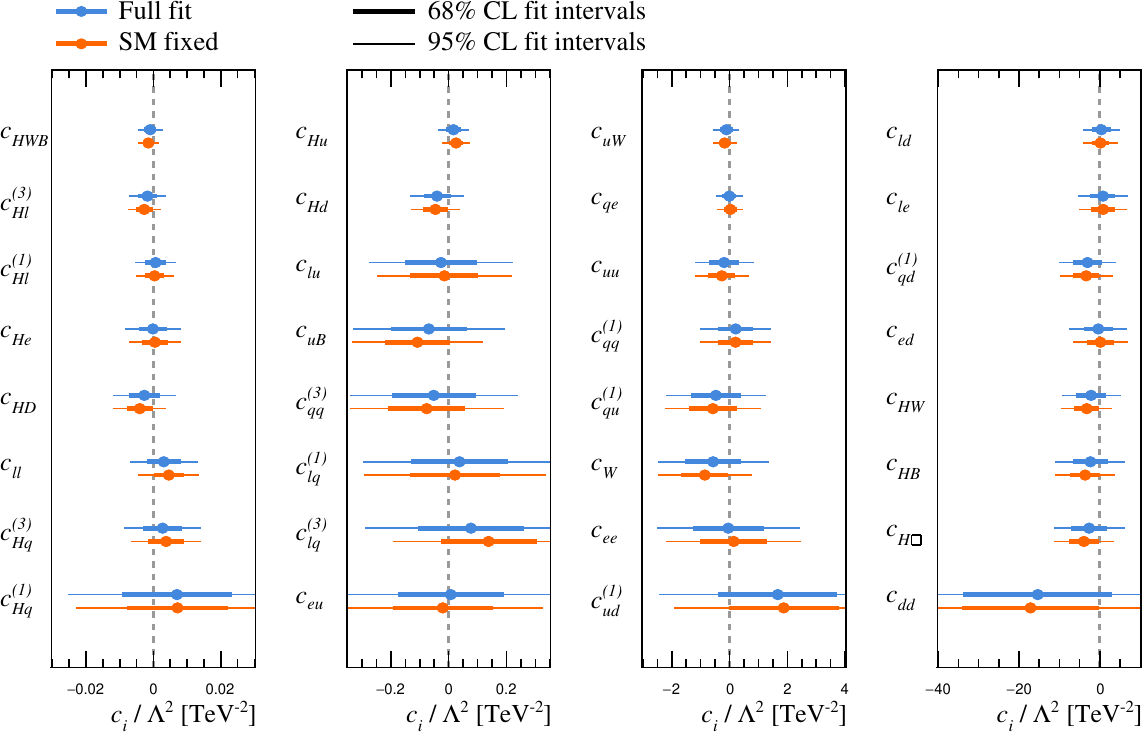}
\vspace{\captionverticalgap}
\caption[]{Summary of the SMEFT bounds on $c_i/\Lambda^2$, shown as 68\% and 95\% CL
intervals from single-coefficient fits. The full fit, profiling the SM inputs and
theory uncertainties together with the Wilson coefficients, is shown in blue, while 
the fixed-SM configuration is shown in orange. \label{fig:smeft_summary}}
\vspace{0.6cm}
\centering
\includegraphics[width=\textwidth]{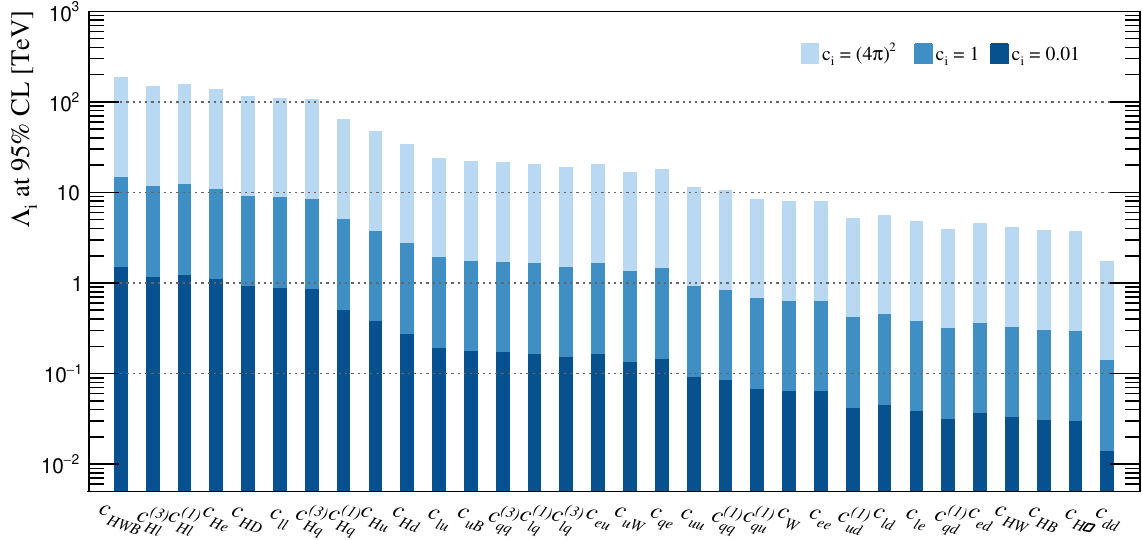}
\vspace{-0.3cm}
\caption[]{Lower bounds at 95\% CL on the characteristic BSM scale $\Lambda$ for
$c_i=(4\pi)^2,\;1$, and $0.01$, derived from the single-coefficient fit constraints on 
$c_i/\Lambda^2$ in Table~\ref{tab:smeft} and Fig.~\ref{fig:smeft_summary}. For asymmetric 
intervals, the less restrictive constraint is used. \label{fig:smeft_LambdaLimits}}
\end{figure}

We constrain the ratios $c_i/\Lambda^2$ using the EWPO introduced in
Section~\ref{sec:ewfit} and the SMEFT predictions of
Eq.~\eqref{eq:smeft_shift}. Since the present data do not resolve all operators
simultaneously, we perform single-coefficient fits, varying one coefficient at a
time while setting the others to zero. The resulting 95\% CL intervals are
listed in Table~\ref{tab:smeft} and shown in Fig.~\ref{fig:smeft_summary}. In
the full fit, the SM inputs and theory-uncertainty nuisance parameters are
profiled together with the parameter of interest. In the comparison fit, the SM
predictions are kept fixed and only the corresponding $c_i/\Lambda^2$ ratio is
varied. Figure~\ref{fig:smeft_LambdaLimits} shows the corresponding lower bounds
on the characteristic BSM scale $\Lambda$ for each dimension-six operator
considered, assuming the benchmark values $c_i=(4\pi)^2$, 1, and 0.01 for 
the Wilson coefficients.

The bounds on $c_i/\Lambda^2$ span more than four orders of magnitude. The strongest 
constraints, $\Order(10^{-3})\tev^{-2}$, are obtained for operators entering the gauge
boson propagators or leptonic $Z$ couplings at tree level: $c_{HWB}$ and $c_{HD}$,
which generate $S$ and $T$, together with $c_{H\ell}^{(1)}$, $c_{H\ell}^{(3)}$,
$c_{He}$, and the four-lepton operator $c_{\ell\ell}$ entering the extraction of \GF.
For $c_i\simeq1$, these correspond to scales of order
$\Lambda\simeq10$--$15\tev$. Operators contributing only at NLO are much less
constrained. Purely four-quark operators, such as $c_{qq}^{(1,3)}$ and $c_{uu}$,
are bounded at $\Order(1)\tev^{-2}$. Their sensitivity arises entirely from the
NLO QCD and electroweak corrections to the $Z$-pole and $W$ observables
computed in Ref.~\cite{Dawson:2019clf}. The weakest bound of $\Order(50)\tev^{-2}$
is on $c_{dd}$ because it lacks the $m_t^2$-enhanced gauge-boson self-energy 
contribution present for $c_{uu}$.

Profiling the SM inputs and theory uncertainties relaxes the intervals by
typically $10$--$15\%$, and at most about $20\%$, while leaving the central values
and hierarchy essentially unchanged. The largest effects occur for the
loop-induced bosonic operators $\mathcal{O}_{HW}$, $\mathcal{O}_{HB}$,
$\mathcal{O}_{H\Box}$ and $\mathcal{O}_{W}$, which are more correlated with the
SM inputs. For operators such as $\mathcal{O}_{qq}^{(1)}$ the effect is
negligible. 

The single-coefficient limits agree with the NLO analysis of
Ref.~\cite{Dawson:2019clf}, including the pattern of sensitivities and the
appearance of four-fermion operators only at NLO. Residual differences are due
to the updated inputs, in particular the new \MW and \sinleff{} combinations of
Section~\ref{sec:impr}, and to profiling the SM and theory-uncertainty
parameters. A complementary SMEFT likelihood with emphasis on parametric
uncertainties is given in Ref.~\cite{Mildner:2024wbl}.

A global EWPO-only fit of all contributing operators is underconstrained: at leading 
order ten operators contribute but only eight combinations are probed, while at 
NLO all 32 operators contribute but only ten combinations are constrained, leaving 22
flat directions~\cite{Dawson:2019clf}. 
A physically motivated choice of contributing operators arises only within a specific 
BSM model, whose UV structure generates a definite set of operators. Without such an 
assumption, we restrict the interpretation to single-coefficient bounds and leave 
constraints of sets of operators for future work.

\section{FCC-ee projections}
\label{sec:fccee}

The FCC-ee is expected to substantially improve the precision of the inputs
to the global electroweak and Higgs-coupling fits. We therefore repeat the fits
described above with projected FCC-ee experimental uncertainties and reduced
theoretical uncertainties, and compare the results with the present precision.

\subsection{Fit inputs}
\label{sec:fccee_input}

The projected precision estimates for $M_H$, $M_Z$, $M_W$, $m_t$, $\as(M_Z^2)$, 
$R_\ell^0$, $R_b^0$, $\Gamma_Z$, $\Gamma_W$ and \sinleff at the FCC-ee are taken 
from Ref.~\cite{deBlas:2025gyz,FCC:2025lpp}. For \dalphaHadMZ, we use the uncertainty 
quoted for $\Delta\alpha(M_Z^2)^{-1}$ in the same references, while the projection for
$\sigma_{\rm had}^0$ is taken from Ref.~\cite{deBlas:2019rxi}. Following
Ref.~\cite{deBlas:2025gyz}, theoretical uncertainties affectig the 
predictions and measurements are reduced by a factor of four.

\subsection{Global electroweak fit}
\label{sec:fccee_sm}

\begin{table}[t]
\setlength{\tabcolsep}{0.5pc}
{
\begin{tabular*}{\textwidth}{@{\extracolsep{\fill}}lcccc} 
\hline\noalign{\smallskip}
& \multicolumn{2}{c}{Present}   &   \multicolumn{2}{c}{FCC-ee}  \\
\rs{Parameter} & Indirect & Direct & Indirect & Direct \\
\noalign{\smallskip}\hline\noalign{\smallskip}
$\delta m_\text{t}$ $[\text{GeV}]$ & $\pm1.5$ & $\pm0.6$ & $\pm0.17$ & $\pm0.011$ \\
$\delta M_\text{H}$ $[\text{GeV}]$ & ${}^{+21}_{-18}$ & $\pm0.08$ & $\pm2.3$ & $\pm0.004$ \\
$\delta M_\text{W}$ $[\text{MeV}]$ & $\pm6.1$ & $\pm8.5$ & $\pm1.1$ & $\pm0.24$ \\
$\delta \sin^{2}\theta_\text{eff}^{\ell}$ $(\times 10^{-5})$ & $\pm5.1$ & $\pm16$ & $\pm1.4$ & $\pm0.064$ \\
$\delta \alpha_\text{s}(M_\text{Z}^2)$ $(\times 10^{-2})$ & $\pm28$ & $\pm9$ & \multicolumn{2}{c}{$\pm2.0$} \\
$\delta \Delta\alpha_\text{had}(M_\text{Z}^2)$ $(\times 10^{-5})$ & $\pm32$ & $\pm10$ & $\pm2.7$ & $\pm2.8$ \\
\noalign{\smallskip}\hline\noalign{\smallskip}
\end{tabular*}
}
\caption{Summary of current and FCC-ee projected uncertainties in the global 
electroweak fit. Indirect uncertainties are obtained by removing the corresponding 
direct measurement from the fit. Direct uncertainties denote the experimental input
precision.\label{tab:fccuncert}}
\end{table}

\begin{figure}[t]
{\centering
\begin{tikzpicture}
\node at (0,0) {\includegraphics[width=\defaultDoubleFigureScaleTikz\textwidth]{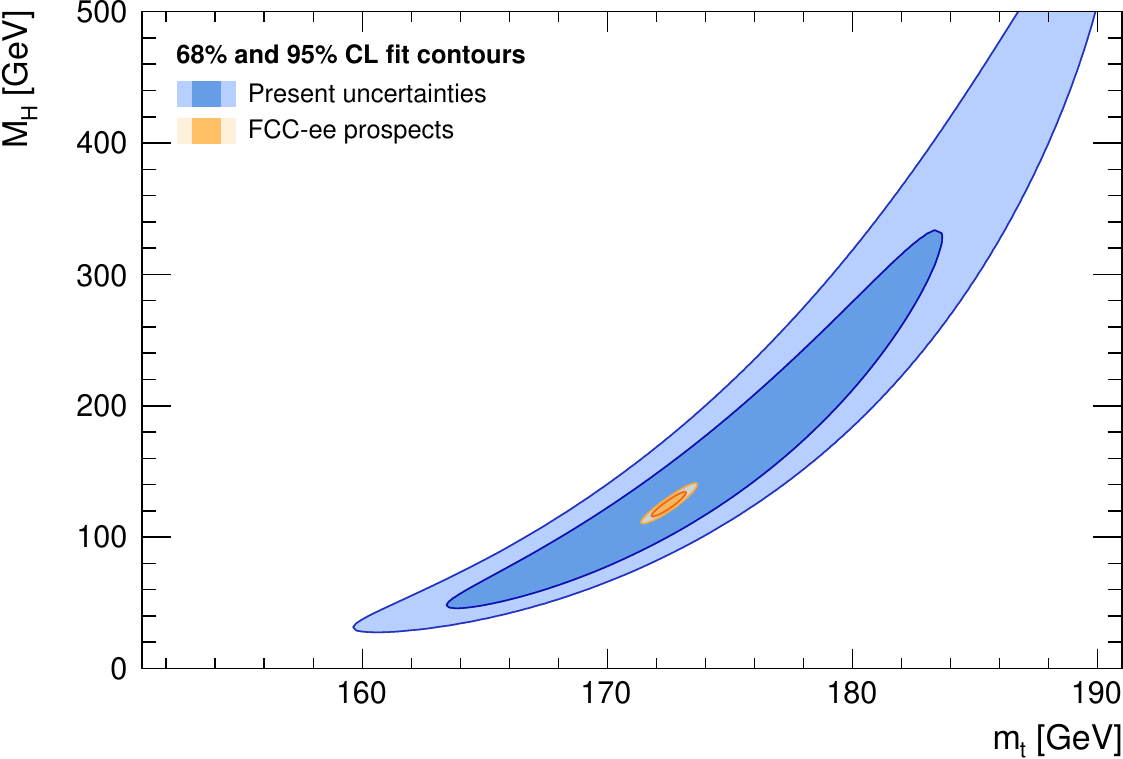}};
\node at (-1.4, 0.4) {\includegraphics[width=.17\textwidth]{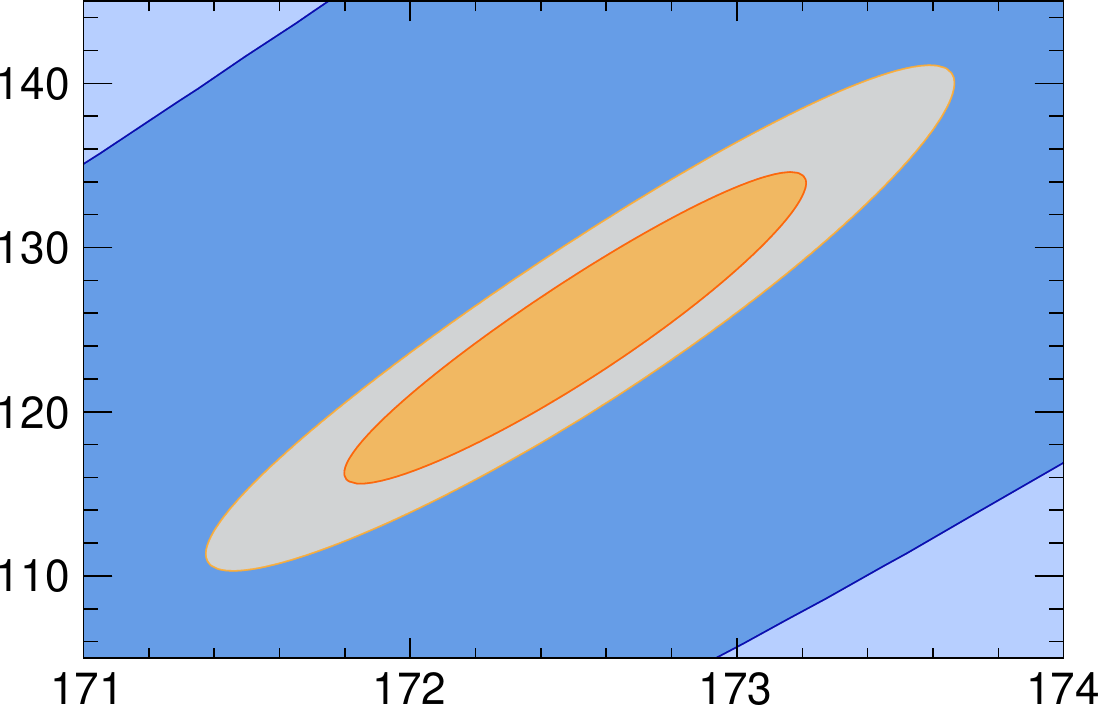}};
\end{tikzpicture}
\begin{tikzpicture}
\node at (0,0) {\includegraphics[width=\defaultDoubleFigureScaleTikz\textwidth]{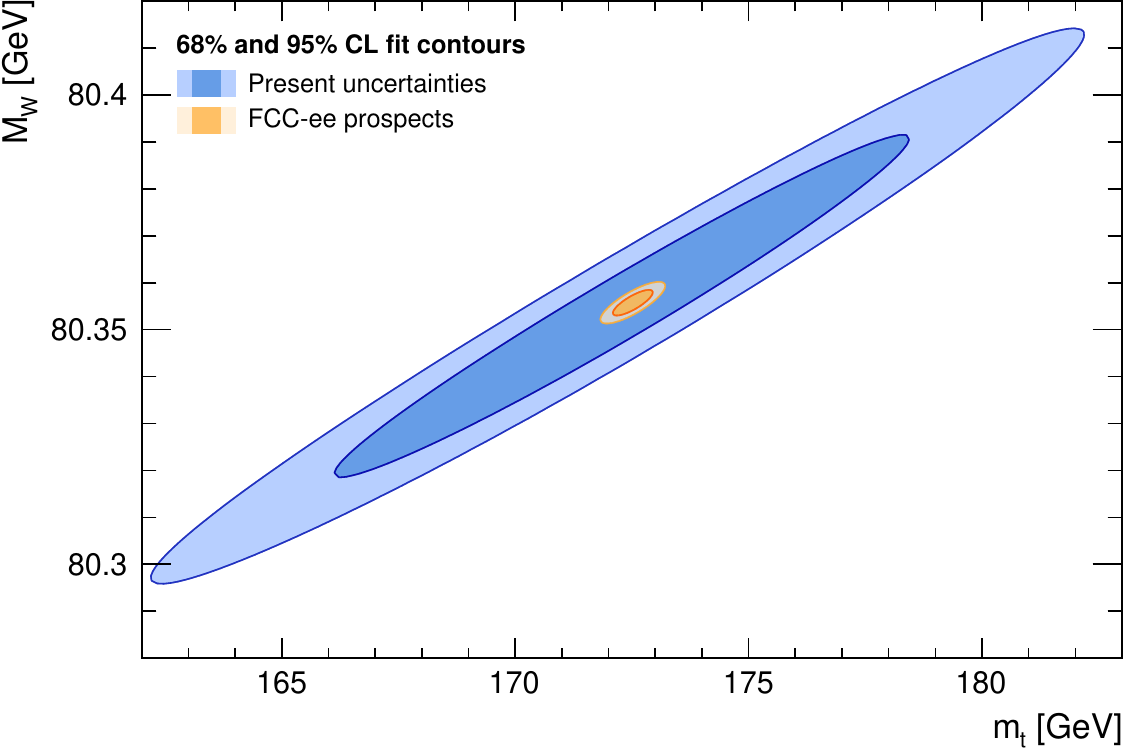}};
\node at (2, -1.0) {\includegraphics[width=.17\textwidth]{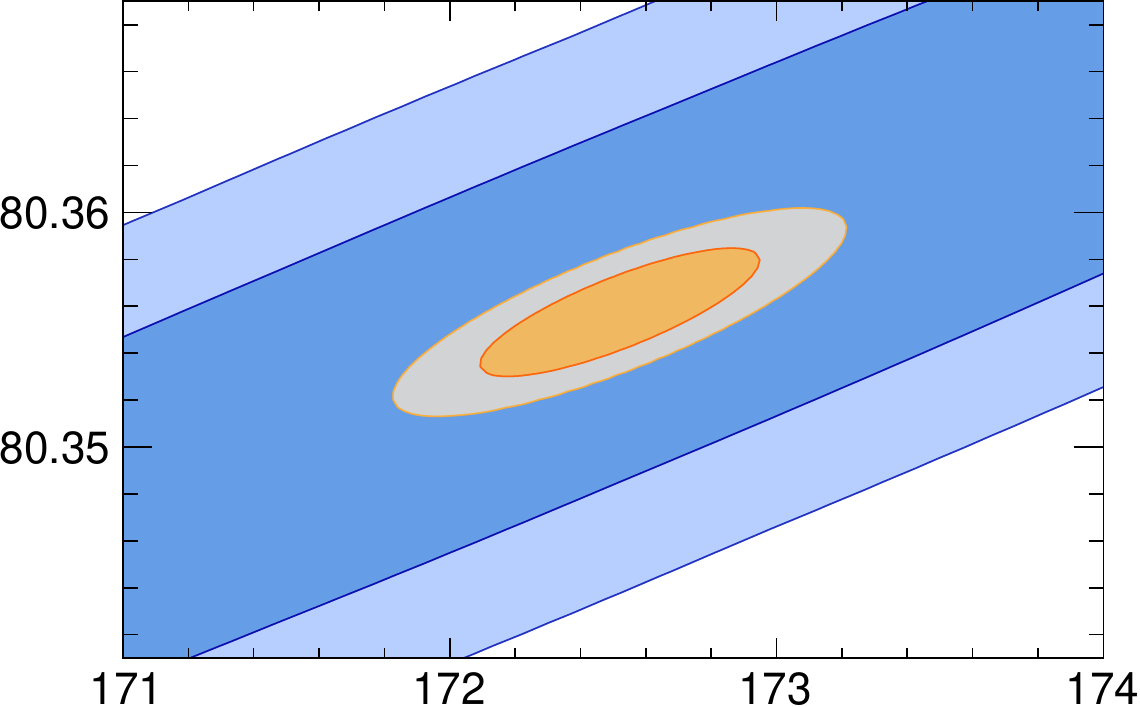}};
\end{tikzpicture} \\
\begin{tikzpicture}
\node at (0,0) {\includegraphics[width=\defaultDoubleFigureScaleTikz\textwidth]{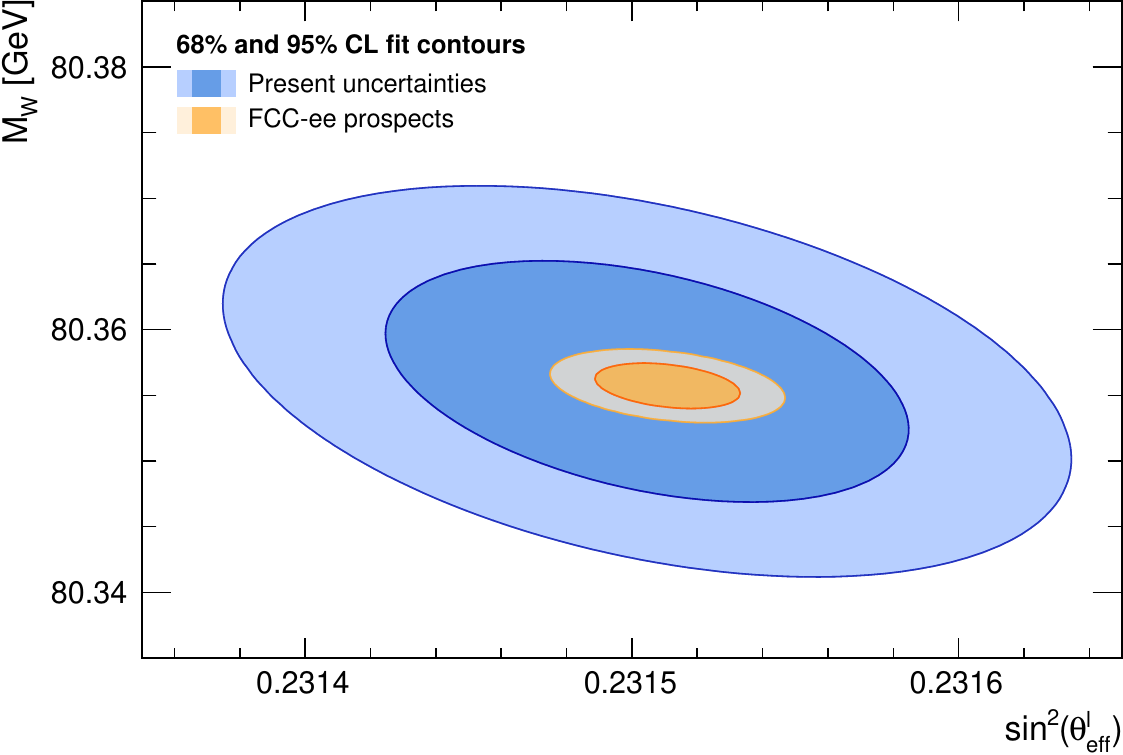}};
\end{tikzpicture}
\begin{tikzpicture}
\node at (0,0) {\includegraphics[width=\defaultDoubleFigureScaleTikz\textwidth]{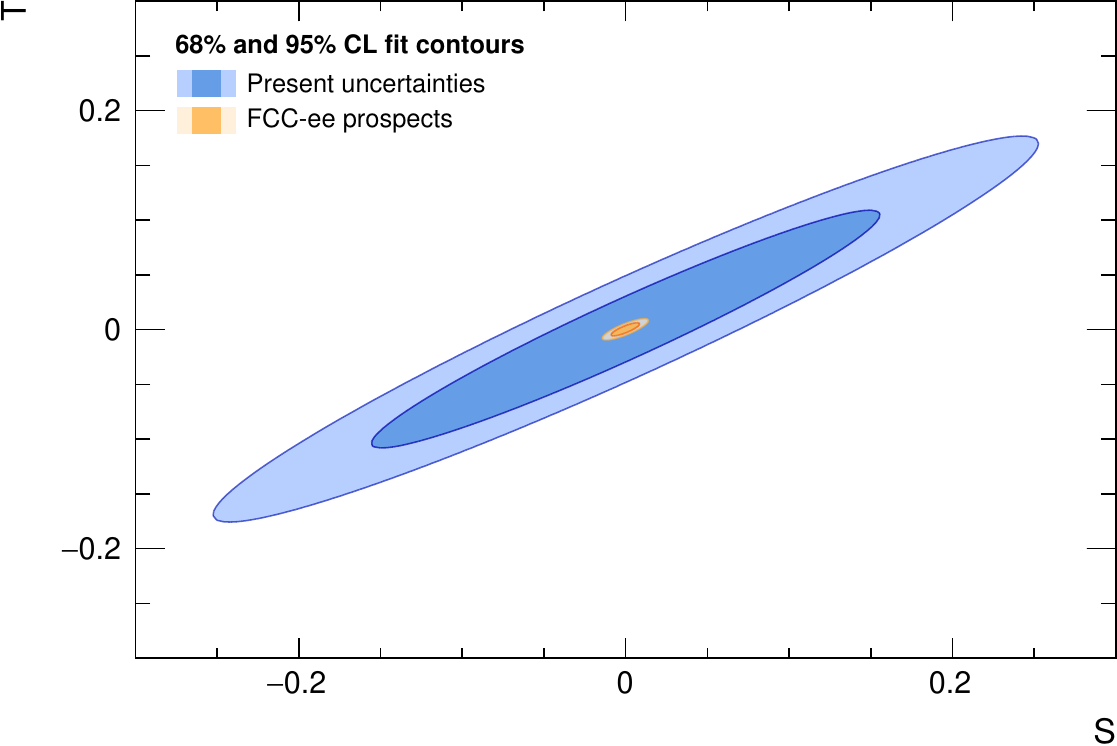}};
\node at (2, -1.0) {\includegraphics[width=.17\textwidth]{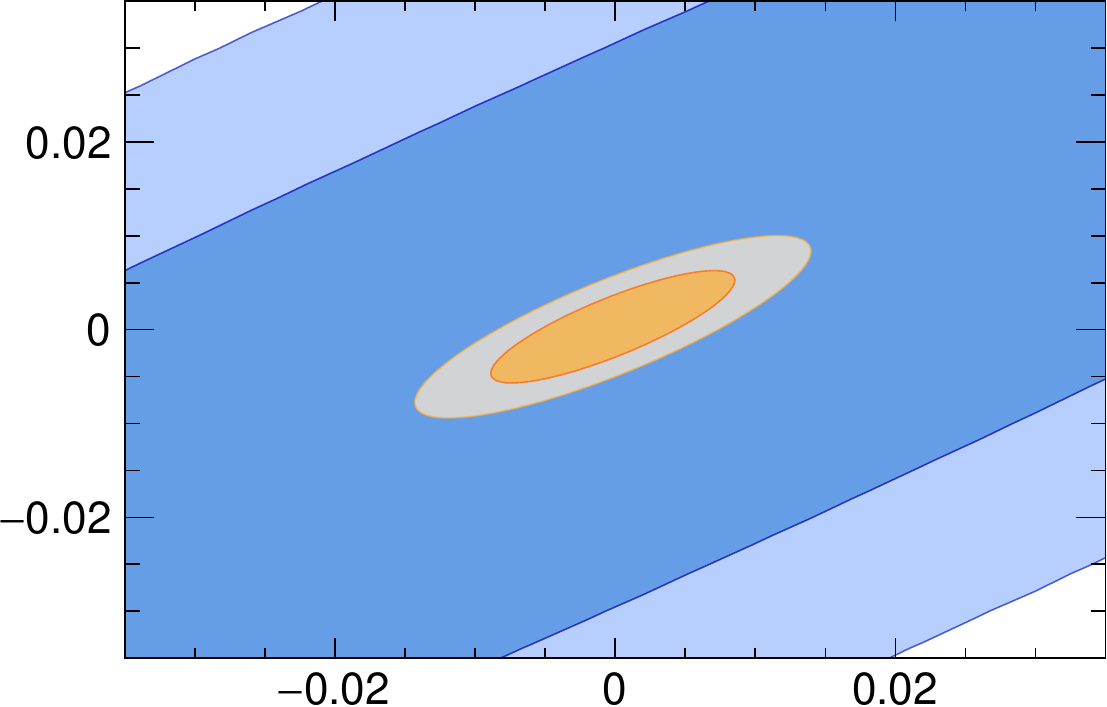}};
\end{tikzpicture}
}
\vspace{-0.4cm}
\caption[]{Two-dimensional scans in the $m_t$--$M_H$, $m_t$--$M_W$, \sinleff--$M_W$, 
and $S$--$T$ planes, comparing the FCC-ee projections with the current global EW fit.
\label{fig:fccee_2d}}
\end{figure}
\begin{figure}[t]
\centering
\includegraphics[width=0.60\textwidth]{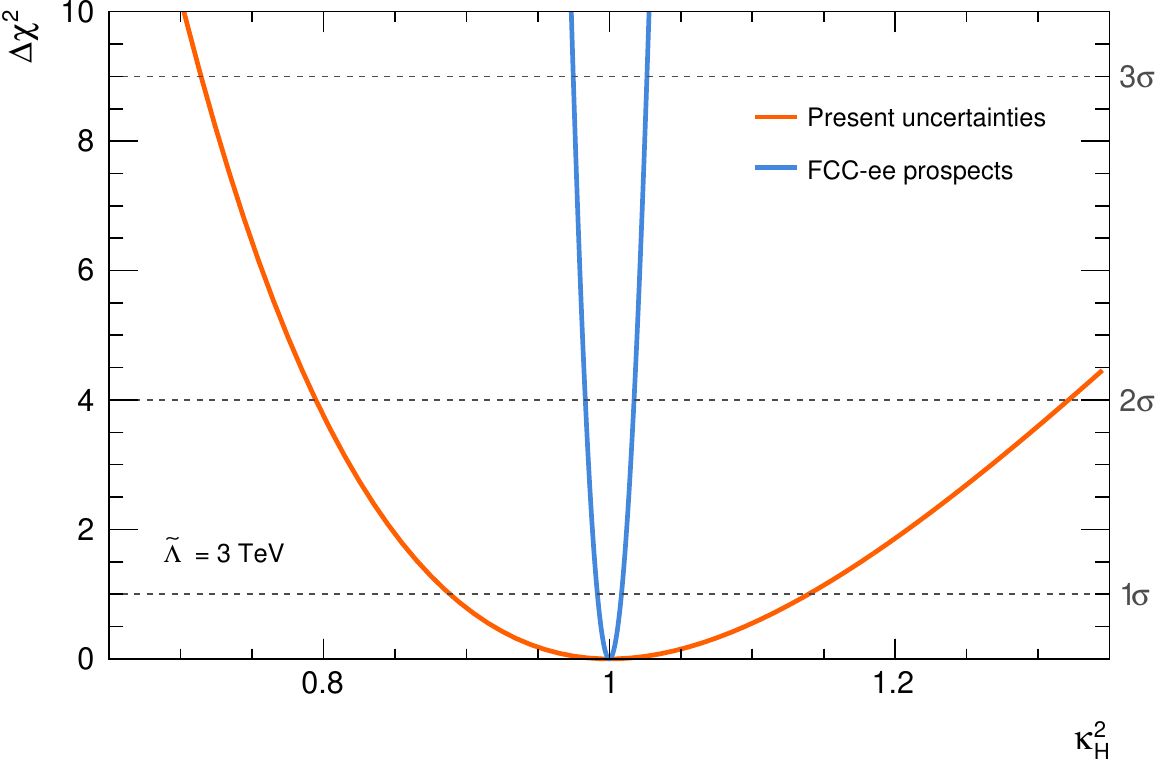}
\vspace{\captionverticalgap}
\caption[]{Scan of $\kappa_H^2$ using projected FCC-ee Higgs signal strengths and EWPO.
\label{fig:fccee_kappaH2}}
\end{figure}

We perform the global electroweak fit with the same configuration as
in Section~\ref{sec:ewfit}, replacing the current input uncertainties by the
FCC-ee projections described above. The central values are chosen such that the
best-fit point has $\chi^2=0$. For comparison, we repeat the fit with the same
central values but with the current uncertainties of Section~\ref{sec:impr}.
Since this projection uses a reduced set of inputs, the corresponding current-fit
uncertainties are very close, but not identical, to those in
Table~\ref{tab:results}.

The projected uncertainties for the main observables are given in
Table~\ref{tab:fccuncert}, together with their expected measurement
precisions. For several observables, the FCC-ee projections improve the indirect
constraints by about an order of magnitude. Figure~\ref{fig:fccee_2d} shows the
corresponding two-dimensional constraints in the $M_H$, $M_W$, $m_t$ and
\sinleff planes, as well as in the $S$--$T$ plane, compared with the current
fit results.

\subsection{Higgs-coupling projections}
\label{sec:fccee_higgs}

The FCC-ee will also provide precise Higgs measurements, in particular through
the Higgs-strahlung process. We use projected signal strengths for
$e^+e^-\to ZH$ with $H\to WW$ and $H\to ZZ$,
$\mu_{ZH}^{WW}=1\pm0.008$ and
$\mu_{ZH}^{ZZ}=1\pm0.025$~\cite{Selvaggi:2025kmd}, together with the projected
EWPO described in Section~\ref{sec:fccee_input}. In the {\em bosonic parameterisation},
these inputs constrain \kappaV and $\kappa_H^2$, with the EWPO providing the
independent \kappaV information through the oblique parameters.

The projected scan of $\kappa_H^2$ is shown in
Fig.~\ref{fig:fccee_kappaH2}. For comparison, we also show the current result
obtained in the same {\em bosonic parameterisation}, setting all signal-strength central
values to $\mu=1$ and using the present ATLAS and CMS uncertainties. With only
the two projected FCC-ee Higgs inputs and the projected EWPO, the expected
$1\sigma$ uncertainty on $\kappa_H^2$ is $0.008$, compared to $0.14$ from the
current bosonic LHC combination, corresponding to an improvement by almost a
factor of 20.

\section{Conclusions}
\label{sec:conclusions}

We have presented an updated global electroweak fit with the Gfitter framework,
using the latest experimental inputs and state-of-the-art two-loop predictions
for electroweak precision observables (EWPO). The fit yields a $p$-value of $0.68$,
compared with $0.23$ in our previous analysis, owing mainly to the new LHC
combinations of $M_W$ and \sinleff, the updated $\sigma_{\rm had}^0$, and the
completed two-loop calculations. The largest tensions remain those in
$A_{\rm FB}^{0,b}$, at $2.3\sigma$, and $A_\ell(\mathrm{SLD})$, at $-2.0\sigma$.

The indirect determinations through the constrained fit are competitive with, 
or more precise than, several direct measurements. We find $M_W = 80.3558 \pm 0.0061\gev$,
$\sinleff = 0.23150 \pm 0.00005$, and $\mt = 173.6 \pm 1.5\gev$, all consistent 
with the corresponding direct measurements. The oblique parameters remain 
SM-compatible, with $S=\SParam$, $T=\TParam$ and $U=\UParam$, closer to the 
SM point and with reduced uncertainties.

Combining the EWPO with the latest ATLAS and CMS Higgs signal-strength
combinations provides an independent constraint on the \HVV coupling through the
oblique parameters, breaking the approximate flat direction between Higgs
couplings and the total width. Treating $\kappa_H^2$ as a free parameter, we
obtain $\GammaHtot = \GammaHHVV$ in the {\em bosonic parameterisation} and
$\GammaHtot = \GammaHEff$ in the {\em effective parameterisation}, where loop-induced
Higgs production and decays are constrained from data. To our knowledge, this is
the first determination of the total Higgs-boson width at about $10\%$ relative
precision without imposing strong assumptions such as $|\kappaV|<1$. The same
framework gives $\BRBSM < \BRBSMgen$ at 95\% CL in the {\em general parameterisation},
without using direct searches for invisible Higgs decays.

We also interpret the EWPO in the SMEFT, deriving 95\% CL single-coefficient
bounds for 32 dimension-six Wilson coefficients. The strongest limits,
$\Order(10^{-3})\tev^{-2}$, are obtained for operators modifying gauge-boson
propagators or $Z$ couplings at tree level, corresponding to
$\Lambda\simeq15$--$20\tev$ for coefficients of order unity. Operators entering
only at next-to-leading order, including four-quark operators, are less strongly
constrained. Profiling the SM inputs and theoretical uncertainties relaxes the
bounds by about $10$--$15\%$.

Finally, FCC-ee projections illustrate the impact of future high-precision
electron--positron data. The projected global electroweak fit improves several
indirect constraints by up to an order of magnitude, while Higgs-strahlung
measurements combined with EWPO improve the determination of $\kappa_H^2$ by
almost a factor of 20. These results demonstrate the substantially increased 
reach of precision electroweak and Higgs measurements in testing the SM and 
constraining BSM physics.

\subsubsection*{Acknowledgements}
\label{sec:Acknowledgments}

We are grateful to Bogdan Malaescu and Georg Weiglein for useful discussions. 
We acknowledge support from the Deutsche Forschungsgemeinschaft (DFG, German 
Research Foundation) under Germany's Excellence Strategy -- EXC 2121 
``Quantum Universe'' -- 390833306.





%
%
\addcontentsline{toc}{section}{References}
\bibliography{References}{}

\end{document}